\documentclass[lettersize,journal]{IEEEtran}
\usepackage[compact]{titlesec}
\usepackage{amsmath,amsfonts}
\usepackage{algpseudocode}
\usepackage{algorithm}
\usepackage{array}
\usepackage[caption=false,font=footnotesize,labelfont=rm,textfont=rm]{subfig}
\usepackage{textcomp}
\usepackage{stfloats}
\usepackage{url}
\usepackage{verbatim}
\usepackage{graphicx}
\usepackage{cite}
\usepackage{multirow}
\usepackage{multicol}
\usepackage{booktabs}
\usepackage{paralist}
\newcommand{\xxx}{RRTO}
\newcommand{\myparagraph}[1]{\smallskip\textbf{#1.}}
\usepackage{makecell}
\usepackage{xcolor}
\newcommand{\revison}[1]{\textcolor{black}{#1}}

\usepackage{pifont} 
\usepackage{caption}
\newcommand{\cmark}{\ding{51}} 
\newcommand{\xmark}{\ding{55}} 
\newcommand{\github}{\url{https://github.com/hku-systems/RRTO}}

\begin{document}

\title{\xxx: A High-Performance Transparent Offloading System for Model Inference \revison{in Mobile Edge Computing}}


\author{Zekai~Sun, Xiuxian~Guan, Zheng~Lin, Yuhao~Qing, Haoze~Song, Zihan~Fang, Zhe~Chen,~\IEEEmembership{Member,~IEEE,} Fangming~Liu,~\IEEEmembership{Senior Member,~IEEE,} Heming~Cui,~\IEEEmembership{Member,~IEEE}, Wei~Ni,~\IEEEmembership{Fellow,~IEEE}, Jun~Luo,~\IEEEmembership{Fellow,~IEEE}  
\thanks{Z. Sun, X. Guan, Y. Qing, H. Song, and H. Cui are with the Department of Computer Science, University of Hong Kong, Pok Fu Lam, Hong Kong SAR, China.  Z. Sun, Y. Qing and H. Cui are also with Shanghai AI Laboratory, Shanghai, China. (e-mail: zksun@cs.hku.hk; xxguan@cs.hku.hk; yhqing@cs.hku.hk; hzsong@cs.hku.hk; heming@cs.hku.hk).}
\thanks{Z. Lin is with the Department of Electrical and Electronic Engineering, University of Hong Kong, Pok Fu Lam, Hong Kong SAR, China (e-mail: linzheng@eee.hku.hk).}
\thanks{Z. Fang is with the Department of Computer Science, City University of Hong Kong, Kowloon, Hong Kong SAR, China (e-mail: zihanfang3-c@my.cityu.edu.hk).}
\thanks{Z. Chen is with the Institute of Space Internet, Fudan University, Shanghai 200438, China, and also
 with the School of Computer Science, Fudan University, Shanghai 200438,
 China (e-mail: zhechen@fudan.edu.cn).}
\thanks{F. Liu is with the Peng Cheng Laboratory, Shenzhen, China, and Huazhong University of Science and Technology, Wuhan, China (e-mail: fmliu@hust.edu.cn).}
\thanks{W. Ni is with Data61, CSIRO, Marsfield, NSW 2122, Australia, and the School of Computing Science and Engineering, and the University of New
 South Wales, Kennington, NSW 2052, Australia (e-mail: wei.ni@ieee.org).}
\thanks{J. Luo is with the School of Computer Engineering, Nanyang Technological University, Singapore (e-mail: junluo@ntu.edu.sg).}

}

\markboth{}%
{Shell \MakeLowercase{\textit{et al.}}: A Sample Article Using IEEEtran.cls for IEEE Journals}


\maketitle

\begin{abstract}
\revison{Deploying Machine Learning (ML) applications on resource-constrained mobile devices remains challenging due to limited computational resources and poor platform compatibility.
While Mobile Edge Computing (MEC) offers offloading-based inference paradigm using GPU servers, existing approaches are divided into non-transparent and transparent methods, with the latter necessitating modifications to the source code.
Non-transparent offloading achieves high performance but requires intrusive code modification, limiting compatibility with diverse applications. 
Transparent offloading, in contrast, offers wide compatibility but introduces significant transmission delays due to per-operator remote procedure calls (RPCs). 
To overcome this limitation, we propose \xxx{}, the first high-performance transparent offloading system tailored for MEC inference. 
\xxx{} introduces a record/replay mechanism that leverages the static operator sequence in ML models to eliminate repetitive RPCs. 
To reliably identify this sequence, \xxx{} integrates a novel Operator Sequence Search algorithm that detects repeated patterns, filters initialization noise, and accelerates matching via a two-level strategy. 
Evaluation demonstrates that \xxx{} achieves substantial reductions of up to 98\% in both per-inference latency and energy consumption compared to state-of-the-art transparent methods and yields results comparable to non-transparent approaches, all without necessitating any source code modification.}
\end{abstract}

\begin{IEEEkeywords}
Computation Offloading, model inference, \revison{mobile edge computing}, distributed system and network
\end{IEEEkeywords}

\section{Introduction}

Machine learning (ML) has become fundamental to a wide range of mobile applications, from intelligent wearable sensors~\cite{luo2021binarized,tang2024merit} and autonomous vehicles~\cite{saridena2022dnn,lin2022channel,fang2024ic3m} to industrial IoT systems~\cite{lin2024fedsn,zhao2019novel,peng2025sigchord,yuan2025constructing,chen2021rf,lin2021spatial,zhang2024fedac,yuan2024satsense,peng2024sums,zhao2024leo}.
These applications are built upon advances in object detection~\cite{lin2025hierarchical,kapao,lin2024efficient}, robotic control~\cite{agrnav,duan2025rethinking,lin2023pushing,zhang2025state,lin2025hasfl}, and environmental perception~\cite{cao2022monoscene}, all of which require high-performance (low-latency and energy-efficient) inference.
\revison{
Deploying ML models on real-world mobile devices (e.g., smartphones, robots, and IoT devices) faces two main challenges: platform compatibility, which requires supporting diverse software frameworks and hardware accelerators on mobile devices; and limited on-device computing resources~\cite{lyu2023optimal,lin2024splitlora,sun2025intra,lin2024adaptsfl}, including restricted computational power and battery life.
}

\revison{
Recent studies~\cite{abbas2017mobile} have proposed mobile edge computing (MEC) as a promising solution for high-performance inference by leveraging GPU servers (e.g., edge devices with powerful GPUs) at the edge of the radio access network. 
Conventional MEC approaches (illustrated in Tab.~\ref{tab:method-comparison}) fall into three categories: device-only inference, non-transparent offloading, and transparent offloading, based on whether source code modification is required for enabling offloading.
Device-only inference demands significant engineering effort due to poor platform compatibility and cannot deliver high performance because of limited on-device resources (see Sec.~\ref{sec:device}). 
As a result, many ML applications are turning to offloading-based inference over MEC networks, which offers high performance and broad compatibility by leveraging GPU servers.}

\begin{table*}[!t]

\setlength\abovecaptionskip{6pt}
\setlength\belowcaptionskip{-15pt}
    \setlength{\tabcolsep}{5pt}
    \centering
    \small
    \color{black}
    \begin{tabular}{lccccc}
        \toprule
        Method 
        & Category
        & Code Modification
        & Application Diversity
        & Platform Compatibility
        & High Performance \\
        \midrule
        Device-only Inference   & Device-only       & N/A   & \cmark & \xmark & \xmark \\
        Native Offloading       & Non-transparent   & High  & \xmark & \cmark & \cmark \\
        CUDA Graph~\cite{qiao2020best}        & Non-transparent   & Low   & \xmark & \xmark & \cmark \\
        Cricket~\cite{cricket}                 & Transparent       & N/A   & \cmark & \cmark & \xmark \\
        \midrule
        \textbf{\xxx{} (Ours)}           & Transparent       & N/A   & \cmark & \cmark & \cmark \\
        \bottomrule
    \end{tabular}
    \caption{ \small 
        \textbf{\revison{Comparison of representative inference methods in MEC.}}
    }
    \label{tab:method-comparison}
\end{table*}

Non-transparent offloading strategies enhance model inference performance by relocating model computations to GPU servers by modifying applications' source code. 
For instance, our experiments demonstrate that native offloading, where the entire model is hosted remotely, achieves up to 3.7 times faster inference and 49\% lower energy consumption compared than device-only inference. 
\revison{
However, this inherent requirement for source code modification poses a critical limitation, which not only significantly increases engineering overhead but also severely curtails application diversity (i.e., support various upper-layer applications). 
Specifically, it impedes integration with closed-source software (e.g., TensorRT~\cite{davoodi2019tensorrt} and CUDA-X libraries~\cite{yi2024study}) and runtime-optimized model structures (e.g., Just-In-Time compilation~\cite{mounesan2025infer}) where code modifications are often infeasible or prohibitive.
While alternatives like CUDA Graph~\cite{qiao2020best} and TorchScript~\cite{devito2022torchscript} have attempted to mitigate the extent of these modifications via model-level packaging, they remain intrinsically non-transparent and suffer from limited platform compatibility due to their framework-specific nature (see Sec.~\ref{sec:non-transparent}).}

Transparent offloading methods~\cite{cricket} enable model inference to be offloaded to GPU servers without code modification, offering convenience at the cost of reduced performance. 
During inference, ML models execute a sequence of operators (e.g., \textit{torch.nn.functional.add()}, \textit{torch.nn.functional.conv2d()} in PyTorch~\cite{pytorch}), which are typically forwarded to backend system functions (e.g., \textit{aten::add} and \textit{aten::conv2d} within CUDA’s ``cudaLaunchKernel''~\cite{cudaruntime}). 
Transparent offloading intercepts these calls and redirects their execution to GPU servers via Remote Procedure Calls (RPC), avoiding source code changes (see Sec.~\ref{sec:transparent}). 
However, since each operator triggers a separate RPC, the accumulated Round-Trip Time (RTT) introduces significant transmission overhead in MEC networks.
This is because ML models often contain hundreds of operators (e.g., 522 in~\cite{kapao}) and invoke thousands of RPCs during inference (e.g., 5895 in~\cite{kapao}); in wireless networks commonly used for mobile devices, each RTT can take several milliseconds~\cite{schneider2022evaluation}, making transmission delay dominate inference time (up to 95\% for ~\cite{cricket}).

\revison{
To address these challenges, one idea is to reduce transmission delay in transparent offloading by shifting from a reactive to a preactive approach. 
Existing transparent offloading methods trigger RPCs only after operators are invoked during inference, leading to sequential scheduling and communication delays. 
This reactive behavior stems from their general-purpose design for remote GPU usage, where operator patterns in upper-layer applications are often unpredictable, making proactive scheduling infeasible. 
In contrast, ML inference typically follows a fixed and predictable operator sequence due to the static computation graph (see Sec.~\ref{sec:observation}). 
This enables a preactive offloading mechanism that uses operator tracing and record/replay to anticipate operator sequences. 
By directly replaying recorded operators on the GPU server, the preactive approach eliminates per-operator RPC communication during the offloading process, significantly reduces transmission delay, and brings the performance benefits of non-transparent offloading back to transparent offloading.
}

\revison{
Integrating the record/replay mechanism into transparent offloading systems presents three key challenges in accurately identifying the correct operator sequence for each inference. 
First, transparent offloading systems intercept function calls from upper-layer applications to GPU devices at the system layer, where \xxx{} can only access low-level log records of the operators called without any direct hints from upper-layer applications, making it difficult to determine which inference task each operator belongs to. 
Second, operator RPCs generated during model loading and initial inference may differ from those in the regular inference loop, introducing variability that complicates sequence identification, where even small mismatches can compromise correctness. 
Third, large operator traces, often containing tens of thousands of entries, create a vast search space that hinders timely identification of the correct sequence.
}

To tackle the above challenges, we propose \textbf{\xxx}, a \textbf{T}ransparent \textbf{O}ffloading system optimized for model inference in MEC with a novel \textbf{R}ecord/\textbf{R}eplay mechanism: \xxx{} automatically records the operators invoked during the first inferences using traditional RPCs as in traditional transparent offloading systems and replays the identified fixed-order operator sequence in subsequent inferences once the specific operator sequence for each inference is identified. 
\revison{
To accurately identify this sequence, \xxx{} introduces the \textit{Operator Sequence Search} algorithm. 
First, the algorithm leverages the observation that the target sequence is repeatedly embedded in the complete log during multiple inferences, ensuring task-level association of each operator.
Second, the algorithm verifies completeness by checking whether the repeated sequence aligns with the entire log, while ignoring inconsistencies caused by model loading or initialization. 
Third, a three-level fast match algorithm is applied in the Operator Sequence Search algorithm to efficiently reduce the search space and accelerate sequence identification.
The key contributions of this paper are summarized as follows:
\begin{itemize}
    \item To the best of our knowledge, \xxx{} is the first high-performance transparent offloading system tailored for model inference in MEC. 
    The code is released at \github.
    \item We propose a record/replay mechanism for transparent offloading system that eliminates per-operator RPC communication during offloading, significantly reducing transmission delay.
    \item We design the Operator Sequence Search algorithm that accurately reconstructs operator sequences to support the record/replay mechanism.  
    \item We empirically evaluate \xxx{} with extensive experiments. The results demonstrate that \xxx{} outperforms state-of-the-art baselines in MEC without requiring any source code modifications.
\end{itemize}
}

The rest of the paper is organized as follows. 
Sec.~\ref{sec:background} motivates the design of \xxx{} by revealing the challenges in current MEC networks. 
Sec.~\ref{sec:design} presents the system design of \xxx{}. 
Sec.~\ref{sec:implement} introduces the system implementation, followed by performance evaluation in Sec.~\ref{sec:evaluation}. 
Related works and technical limitations are discussed in Sec.~\ref{sec:discussion}.
Finally, conclusions are presented in Sec.~\ref{sec:conclusion}.

\section{Background}
\label{sec:background}
\subsection{\revison{Device-only Inference}}
\label{sec:device}
\revison{
Device-only inference, which runs models directly on mobile devices, faces two major limitations: poor platform compatibility and restricted performance. 
Mobile applications are built on diverse software frameworks (e.g., PyTorch~\cite{pytorch}, TensorFlow~\cite{tensorflow}, CUDA~\cite{cudaruntime}, and OpenCL~\cite{opencl}) and mobile devices are equipped with various hardware accelerators (e.g., GPU~\cite{jia2022codl}, FPGA~\cite{plancher2021accelerating}, and SoC~\cite{lane2016deepx}), making deployment on real-world devices labor-intensive and error-prone. 
Engineers must adapt each model to specific hardware and software configurations, sacrificing portability and increasing cost. 
}

\begin{figure}[htp]
\vspace{-15pt}
\setlength\abovecaptionskip{6pt}
\setlength\belowcaptionskip{-5pt}
\centering
\subfloat[\revison{\textbf{Inference Time}}\label{fig:device-inference}]{\includegraphics[width=0.48\linewidth]{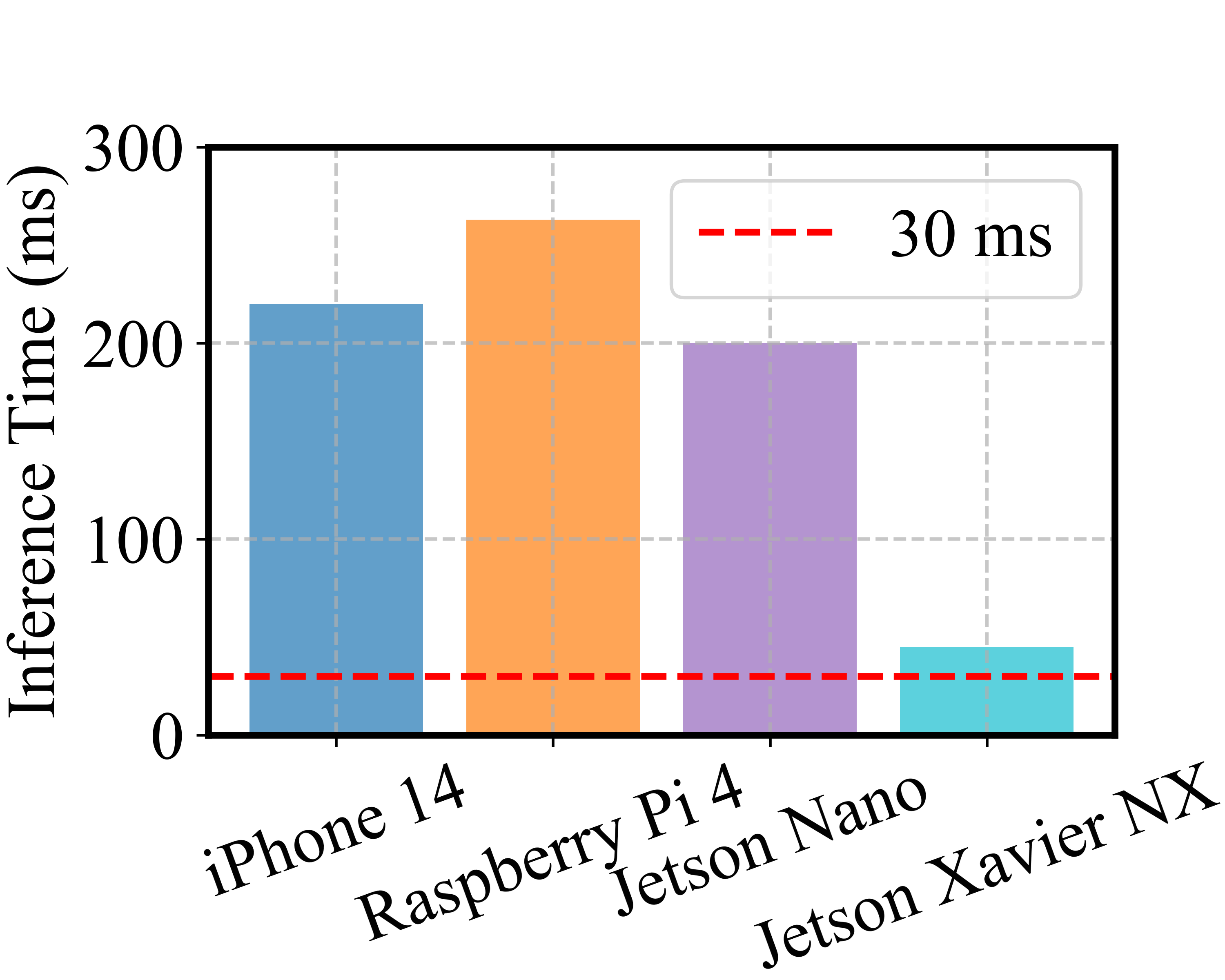}}
\hfil
\subfloat[\revison{\textbf{Battery Life}}\label{fig:device-energy}]{\includegraphics[width=0.48\linewidth]{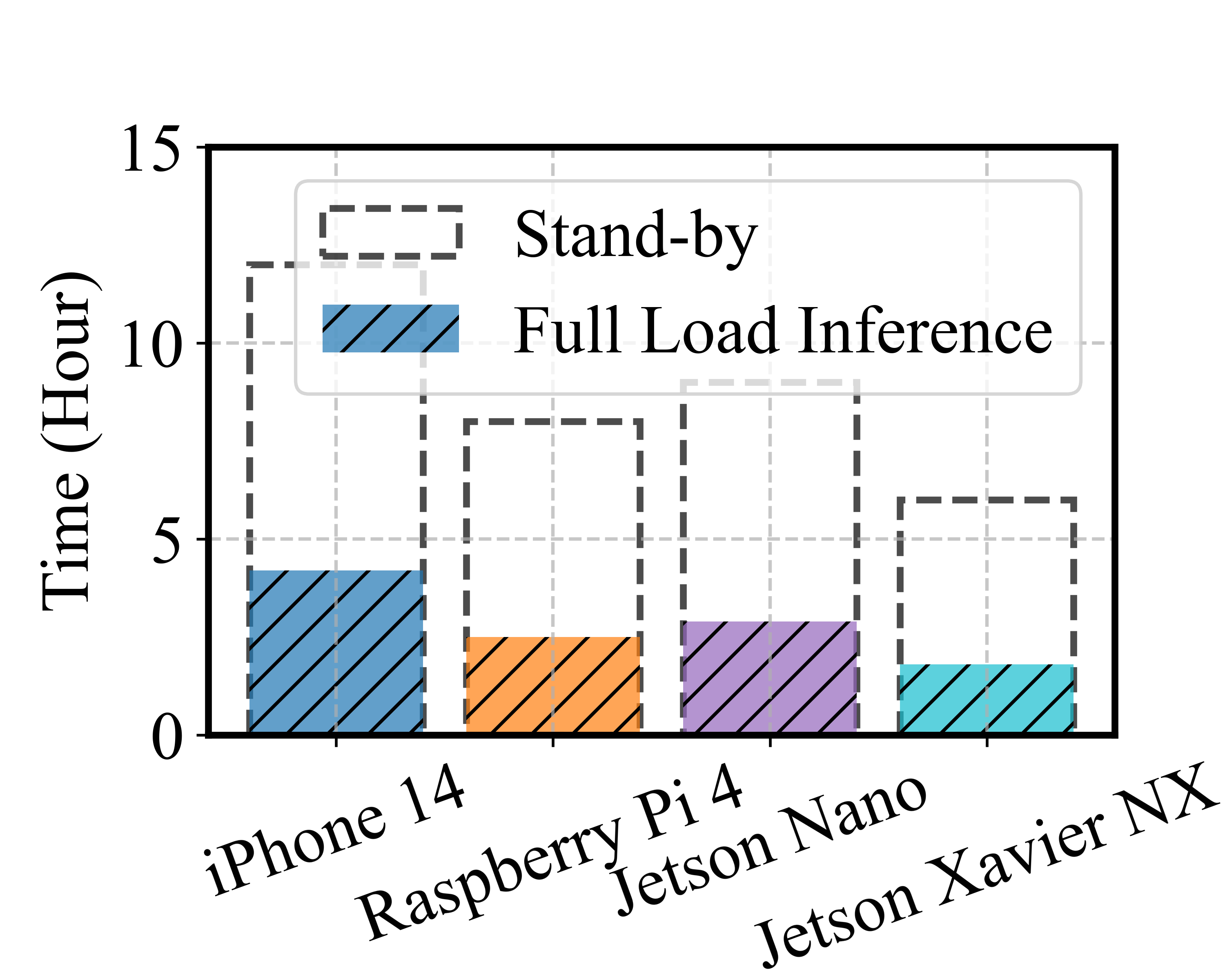}}
\caption{ \small \revison{\textbf{The performance of VGG-16 under device-only inference across different mobile devices~\cite{iphone, raspberrypi, jetsonnx, ning2024power}.}}}
\end{figure}

\revison{
In addition, the performance of device-only inference is constrained by the limited computational power of processors and battery capacity. 
As shown in Fig.~\ref{fig:device-inference}, the inference latency on various mobile devices exceeds the 30 ms threshold required for smooth video fluency~\cite{ghosh2023react} (indicated by the red dotted line).
Fig.~\ref{fig:device-energy} shows that frequent inference reduces device standby time to 20–40\% of their normal duration, degrading user experience and device usability.
}

\subsection{Non-Transparent Offloading}
\label{sec:non-transparent}
Non-transparent offloading strategies relocate model computations to GPU servers by mandating application source code modifications, a requirement that \revison{inherently increases engineering complexity and restricts application diversity. 
This non-transparency severely limits their use, particularly with closed-source software and runtime-adaptive model structures where such code alterations are often infeasible or prohibitive.
High-performance libraries (e.g., TensorRT~\cite{davoodi2019tensorrt} and CUDA-X library~\cite{yi2024study}) are widely adopted in mobile applications but do not expose source code, making them incompatible with non-transparent approaches. 
Many real-world applications also employ runtime optimization techniques, including Just-In-Time compilation~\cite{mounesan2025infer} and dynamic conventional kernel selection with different numbers/sizes based on task-specific requirements~\cite{armanfard2015local}. 
These adaptive methods change model structure at runtime to improve speed and accuracy. 
Non-transparent offloading cannot accommodate such closed-source or dynamic environments. 
In contrast, transparent offloading intercepts system calls during runtime, enabling support for both closed-source libraries and runtime-adaptive model structures without requiring code modification.
}

\revison{
Alternative methods, such as CUDA Graph~\cite{qiao2020best} and TorchScript~\cite{devito2022torchscript},s reduce the amount of code modification by packaging models for GPU server deployment. 
However, these approaches are still non-transparent and cannot support applications involving runtime variability or proprietary libraries. 
Further, they reduce  platform compatibility by depending on specific software frameworks, which imposes greater constraints on mobile software development that already plague device-only deployment today.
}

To further optimize inference latency and energy efficiency, diverse scheduling strategies have been incorporated into non-transparent offloading systems. 
Unlike native offloading, which typically relocates the entire model to GPU servers, these advanced methods employ fine-grained scheduling at various stages of model inference (e.g., layer partitioning~\cite{kang2017neurosurgeon} and multiple inference scheduling~\cite{lin2019computation, fang2017qos, zhao2019novel}). 
The proven effectiveness of these scheduling optimizations within non-transparent frameworks forms a basis for their adaptation to \xxx, with the goal of further enhancing its performance (as detailed in Sec.~\ref{sec:discussion}).


\subsection{Transparent Offloading}
\label{sec:transparent}

\subsubsection{Existing framework}
\begin{figure*}[!t]
\setlength\abovecaptionskip{6pt}
\setlength\belowcaptionskip{0pt}
\centering
\includegraphics[width = 0.85\textwidth]{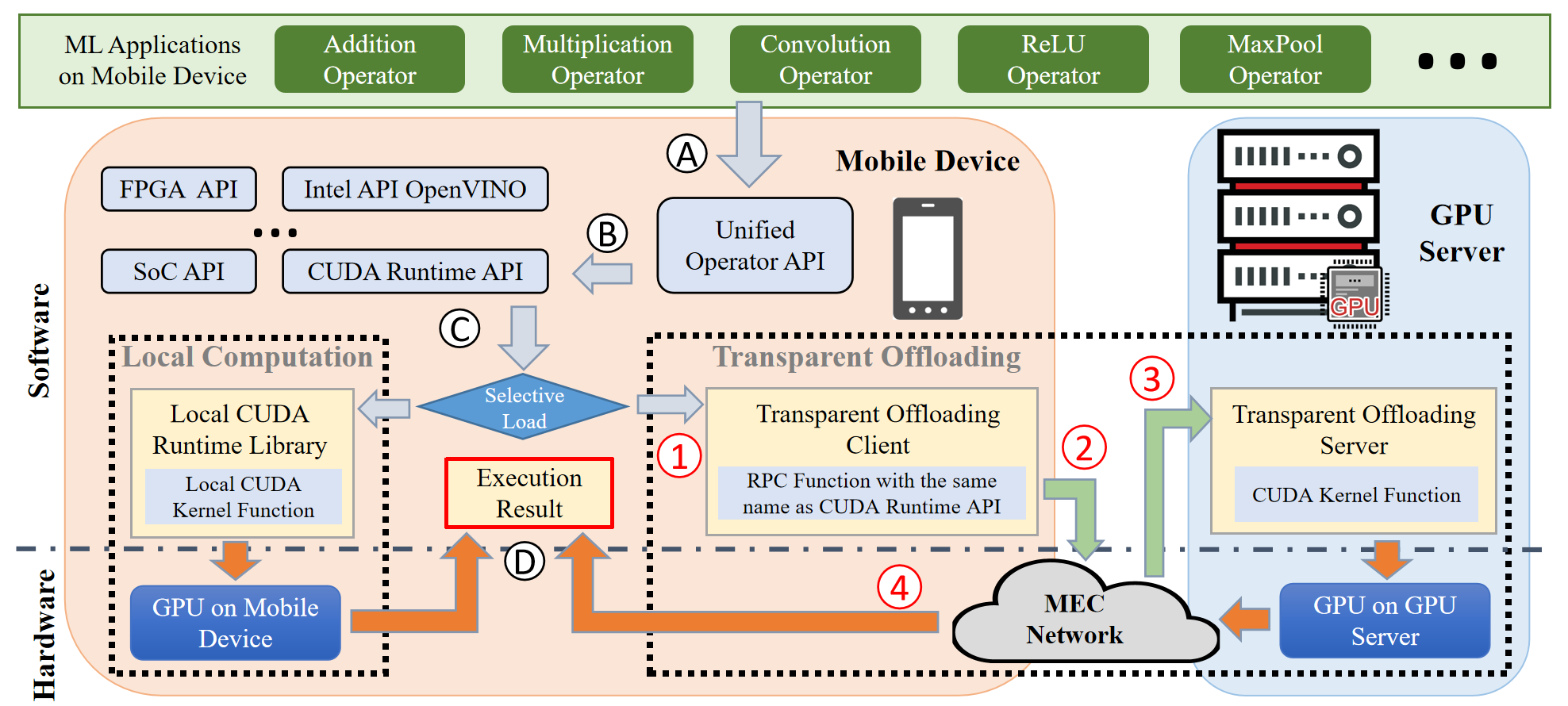}
\caption{ \small \revison{\textbf{Workflow of Transparent Offloading System for Model Inference in MEC.}}}
\label{fig1}
\end{figure*}

When a mobile application utilizes the GPU on the mobile device for inference \revison{(device-only inference)}, the complete, top-down system call flow \revison{for an individual operator} is as follows (illustrated in the left part of Fig.~\ref{fig1}):
\begin{itemize}
  \item [A] 
  The ML application sequentially invokes the corresponding operators based on the ML model's structure to complete the entire computation process.
  \item [B]
  Each operator accesses the appropriate function library through a unified operator API tailored to the type of the device running the application; for instance, on NVIDIA GPUs, this is the CUDA runtime library~\cite{cudaruntime}.
  \item [C]
  The operating system loads the local CUDA runtime library by default, and executes the relevant CUDA kernel functions on the mobile device's GPU.
  \item [D]
  The local CUDA runtime library returns the execution results  \revison{(e.g., `cudaSuccess' from ``cudaLaunchKernel'' and computation result from ``cudaMemcpyDtoH'')} to the upper-layer application.
\end{itemize}

Transparent offloading methods~\cite{cricket} typically employ a strategy of rewriting dynamic link libraries by defining functions that share names with those in the CUDA runtime API and prioritizing the loading of these custom libraries by setting the \textit{LD\_PRELOAD} environment variable. 
This setup causes the dynamic linker to redirect calls from the original library functions to the custom library functions with the same name, effectively intercepting library functions. 
Subsequently, the custom library packages the function calls along with the required parameters and data, and sends them to the GPU server via RPC. 
It also modifies GPU memory management and the launching of CUDA kernel functions to ensure that RPCs are executed correctly on the GPU server. 
By this means, these methods achieve transparent offloading by intercepting kernel functions one-by-one at the system layer using similarly named functions in the dynamic link library and offloading their execution to the GPU server. 

Compared to using the GPU on the mobile device, changes in the system call process primarily occur in step C.
The detailed steps of the transparent offloading process (depicted in the right part in Fig.~\ref{fig1}) are as follows:

\begin{itemize}
  \item [(1)] 
  The dynamic link library is modified such that each operator prioritizes calling the RPC functions that have the same names as those in the CUDA runtime API, enabling the effective identification and interception of all CUDA kernel function calls.
  \item [(2)]
  The transparent offloading client transmits the called CUDA runtime API and the required parameters to the GPU server via the \revison{MEC} network using RPC.
  \item [(3)]
  The transparent offloading server on the GPU server launches the corresponding CUDA kernel functions and completes the computation.
  \item [(4)]
  The execution results are sent back to the client. The transparent offloading system then returns these results to the upper-layer function calls.
\end{itemize}

\subsubsection{Challenges in MEC Networks}
\label{sec:background-transparent}
In real-world scenarios, mobile devices primarily rely on wireless networks, offering high mobility but limited bandwidth compared to data center networks equipped with high-speed technologies (e.g., 200–800 Gbps for InfiniBand~\cite{infiniBand}).

The inherent bandwidth capacity of wireless networks is inherently constrained by both theoretical limits and practical implementation factors in MEC.
While Wi-Fi 6 can achieve a peak bandwidth of 1.2 Gbps per stream~\cite{liu2023first}, mobile devices lack the necessary hardware to fully leverage this capacity~\cite{yang2022mobile}. 
The actual available bandwidth varies significantly due to factors such as device mobility~\cite{masiukiewicz2019throughput}, signal obstruction~\cite{ding2015performance}, and channel contention~\cite{ren2018proportional}.

\begin{figure}[htp]
\vspace{-15pt}
\setlength\abovecaptionskip{6pt}
\setlength\belowcaptionskip{2pt}
    \centering
    \subfloat[\textbf{Indoors}\label{fig:indoors}]{\includegraphics[width=0.5\linewidth]{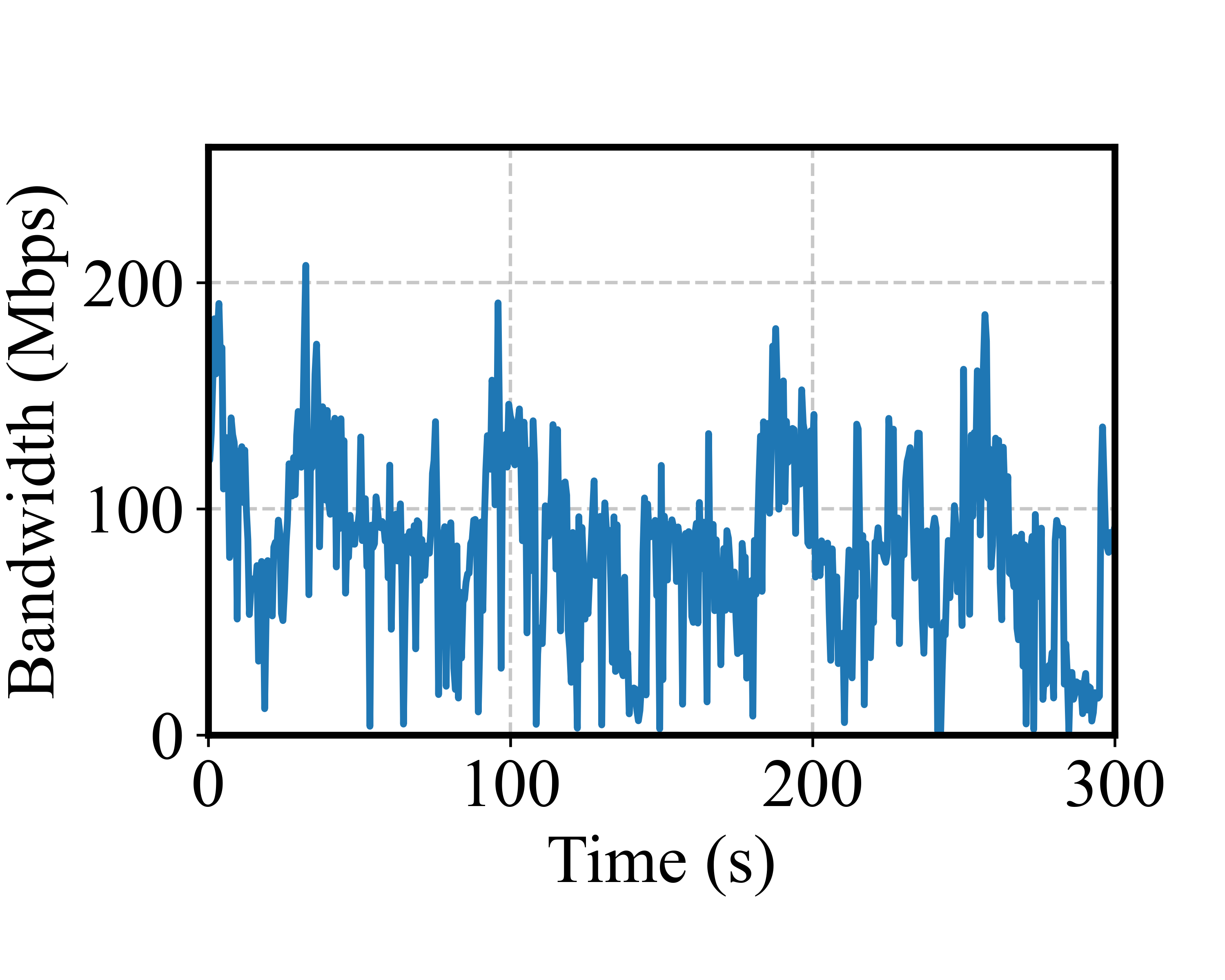}}
    \subfloat[\textbf{Outdoors}\label{fig:outdoors}]{\includegraphics[width=0.5\linewidth]{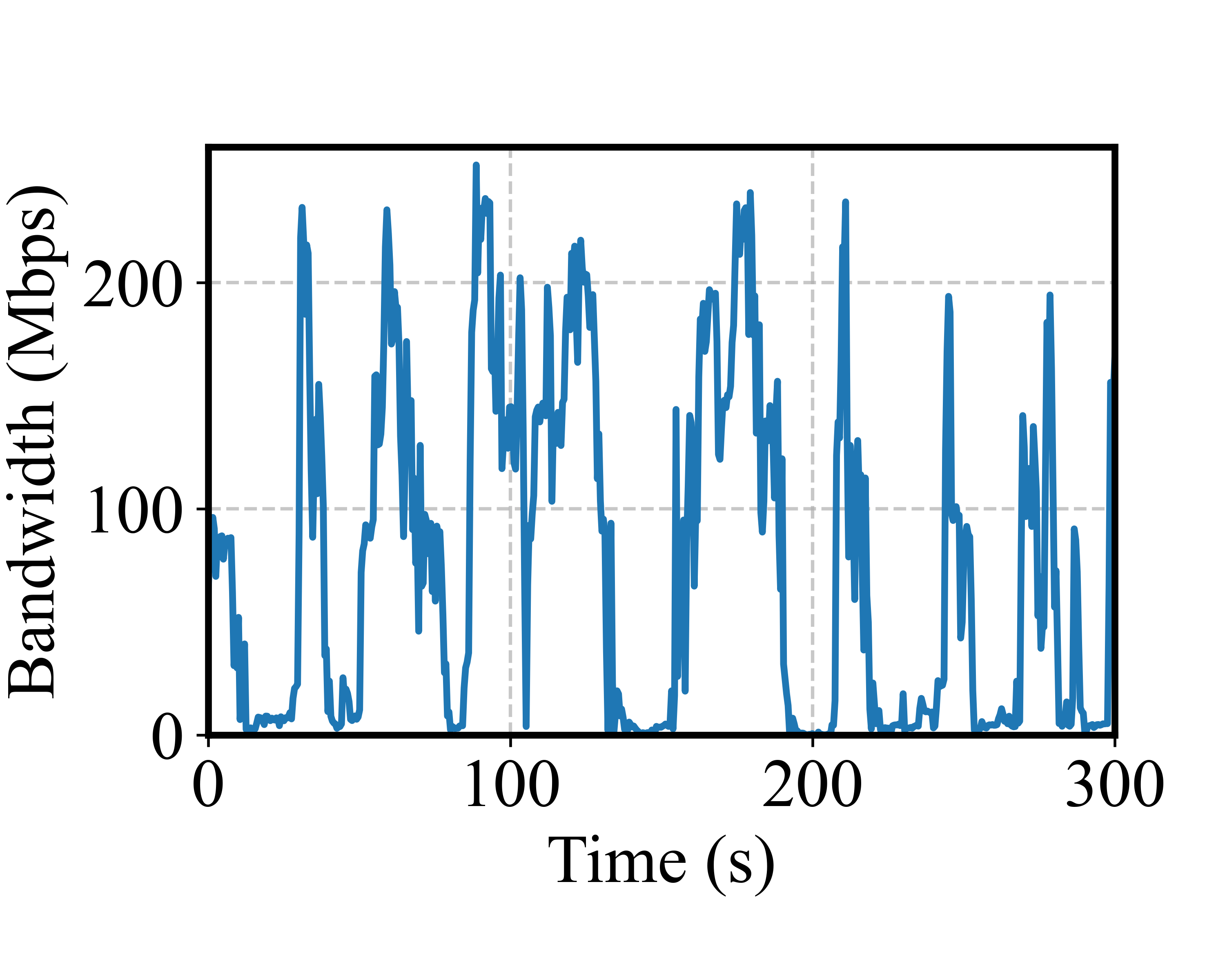}}
    \caption{ \small  \revison{\textbf{The wireless transmission instability of TCP between our robot and the base station in MEC networks.}}}
    \label{fig:bandwidth} 
\end{figure}

To examine wireless instability in MEC scenarios, we conducted a robot surveillance experiment where four-wheeled robots navigated through a lab (indoors) and a campus garden (outdoors) at speeds of 5–40 cm/s. 
Using iperf~\cite{iperf}, we measured real-time wireless bandwidth capacity between the robot and a base station over TCP~\cite{tian2005tcp} at 0.1-second intervals for five minutes. 
As shown in Fig.~\ref{fig:bandwidth}, the average bandwidth was 93 Mbps indoors and 73 Mbps outdoors, with outdoor measurements exhibiting higher fluctuations and occasional near-zero drops due to obstacles and reduced signal reflections.
This indicates that transparent offloading is hard to sustain in real-world wireless networks, as the RTT for operator-level RPCs in traditional transparent offloading systems is usually higher than in data center networks.

\revison{
While the CPU and GPU typically employ an asynchronous paradigm (CUDA kernels are enqueued to the GPU at the inference start, allowing the GPU to process them independently while the CPU awaits overall completion), this paradigm faces severe challenges in MEC due to limited network bandwidth, prolonging the RTT for each remote CUDA kernel launch via RPC. 
This prolonged RTT, often several milliseconds per RPC~\cite{schneider2022evaluation} (and compounded by potentially much longer transfer times for inference inputs/outputs depending on application requirements), starkly contrasts with mere microseconds needed for a local kernel launch command and tens of microseconds to milliseconds for its execution on a GPU server~\cite{nsightcompute}. 
Consequently, the cumulative overhead from frequent RPCs for individual kernel dispatches emerges as a critical performance bottleneck in traditional transparent offloading systems, negating many benefits of the CPU-GPU asynchronous paradigm in constrained network environments.
}

\subsubsection{RPC Optimization}
\label{sec:rpc}
Remote Procedure Call (RPC)~\cite{libtirpc} is a fundamental communication protocol enabling processes to request services from remote computers over a network. 
While common strategies like Caching~\cite{singhvi2021cliquemap} (storing results of previous RPC calls), Batching~\cite{lazarev2021dagger} (aggregating multiple RPC calls into a single request), and Asynchronous RPC~\cite{eyerman2022efficient} (allowing non-blocking execution so the client can perform other tasks while awaiting the server's response) aim to enhance RPC performance, they prove largely ineffective in reducing the substantial communication costs of transparent offloading systems during model inference. 
Specifically, Caching fails because each inference typically processes unique input, necessitating fresh operator computations. 
\revison{
While effective in reducing the number of network requests via aggregation, Batching has a critical drawback: It does not eliminate per-operator RPCs and, more importantly, must rely on latency-inducing timeouts for batch formation. 
This reliance is necessary because the total number of operations is unknown \textit{a priori}, a constraint that our operator search algorithm overcomes.
Furthermore, Asynchronous RPC methods compromise the correctness of inference results because they cannot guarantee the sequential execution of GPU operations on the server. 
This sequential integrity is vital not only for launching CUDA kernels (``cudaLaunchKernel'') in the correct order but for ``cudaMemcpyHtoD'' and ``cudaMemcpyDtoH'' operations to manage input and output data accurately. 
These memory transfer operations, often involving larger data volumes and thus longer transmission times than kernel launches, are particularly prone to misordering under asynchronous execution, leading to corrupted data.}
In contrast, \xxx{} significantly reduces communication costs by eliminating most operator-level RPCs, representing a specialized co-design of RPC optimization and transparent offloading tailored for model inference (see Sec.~\ref{sec:design}).

\section{System Design}
\label{sec:design}
\subsection{Overview}
\label{sec:architecture}

\begin{figure}[htbp]
\setlength\abovecaptionskip{6pt}
\setlength\belowcaptionskip{5pt}
\centering
\includegraphics[width=0.95\linewidth]{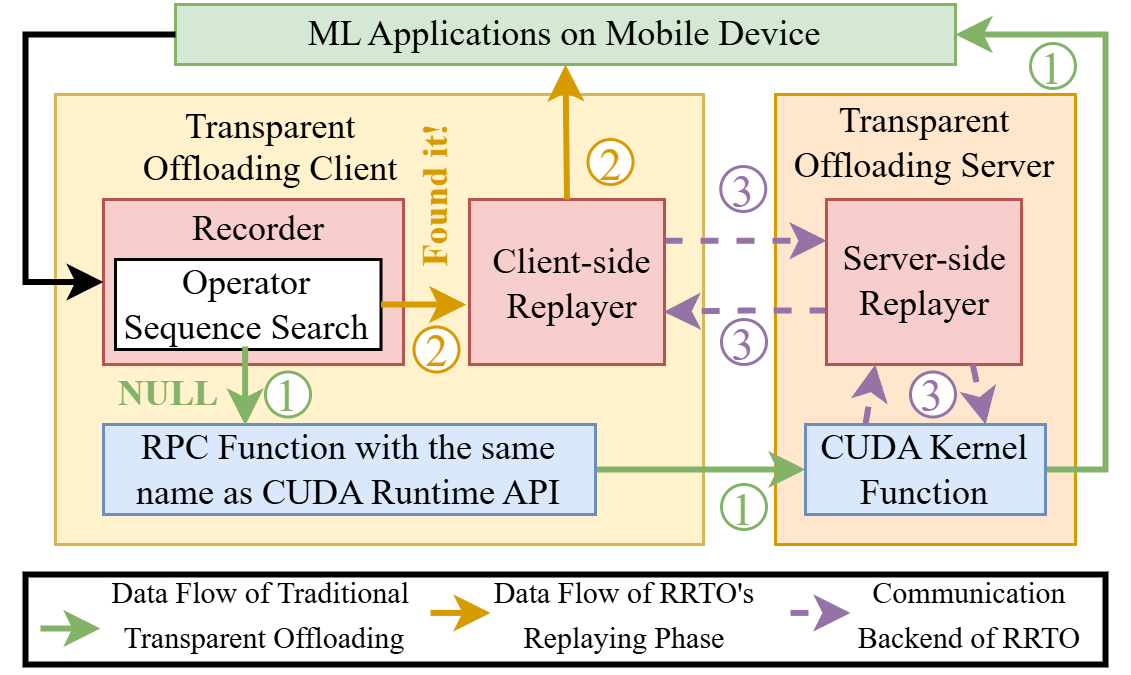}
\caption{ \small \revison{\textbf{Architecture of \xxx, with key components highlighted in red boxes.}}}
\label{fig:architecture}
\end{figure}

This section introduces \xxx, a high-performance transparent offloading system meticulously designed for model inference in MEC, featuring a novel record/replay mechanism. 
The overall architecture of \xxx{} is depicted in Fig.~\ref{fig:architecture}.
Unlike conventional transparent offloading systems (Fig.~\ref{fig1}), \xxx{} adeptly integrates this record/replay mechanism into the core components of the transparent offloading framework while maintaining transparency to upper-layer applications. 
This architectural choice ensures that \xxx{} enables offloading without necessitating any modifications to the application's source code. 
The operational logic for both client and server components is further elaborated on pseudocode in Sec.~\ref{sec:pseudo}.

Within the context of a single inference application, \xxx{} employs its specialized record/replay mechanism to effectively differentiate operators belonging to distinct inference tasks. 
For the first few inference executions, \xxx{} enters \revison{the recording phase (\textbf{\textcircled{1}}).
In this mode, it adheres to the execution pattern of traditional transparent offloading systems, where each operator's execution is individually offloaded to the GPU server via RPCs, as illustrated by the green lines in Fig.~\ref{fig:architecture}.}
When \xxx{} intercepts CUDA kernel function calls originating from upper-layer ML applications, its recorder component logs these invoked functions, including their parameters and return values. 
Concurrently, it performs an operator sequence search to precisely identify the sequence of inference operators (see Sec.~\ref{sec:oss}).

Once the recorder successfully identifies the complete inference operator sequence, \xxx{} transitions to \revison{the replaying phase (\textbf{\textcircled{2}}). 
During this phase, subsequent inferences are executed by replaying the pre-identified operator sequence through replayer components strategically deployed on both the client and server, as illustrated by the orange lines in Fig.~\ref{fig:architecture}. 
The client-side replayer assumes a pivotal role to ensure \xxx's transparency and high-performance simultaneously: It intelligently furnishes the upper-layer applications with the execution results from previously recorded RPC calls (mainly `cudaSuccess' from ``cudaLaunchKernel''), a mechanism analogous to the RPC caching techniques employed in existing optimization methods (see Sec.~\ref{sec:rpc}). 
This approach enables the offloading client seamlessly invokes system functions for subsequent operators, creating the illusion of local execution, until it reaches the end operator. 
Upon encountering this operator, it pauses to await the actual inference output, which is computed and transmitted back from the server-side replayer.
Concurrently with this primary replay operation, the client-side replayer monitors for any deviations or failures in the predicted operator sequence and adeptly detects the initiation of new inference tasks.
In this way, \xxx{} maintains the transparency and system integrity while eliminating per-operator RPC communication overhead during the replaying phase.
}

Simultaneously, the server-side replayer efficiently performs the operator computations on the GPU server \revison{(\textbf{\textcircled{3}}).
It replays the identified sequence and transmits only the final inference result back to the client. 
This innovative approach enables \xxx{} to achieve a communication overhead closely approximating that of non-transparent offloading systems.
Specifically, it transmits only the raw input and final output data, and circumvents per-operator RPC communication during the offloading process in the replaying phase, as visualized by the purple dotted lines in Fig.~\ref{fig:architecture}.}

\revison{
When multiple ML applications run concurrently on a mobile device, \xxx{} distinguishes operators from different applications by leveraging the inherent capability of existing transparent offloading systems to separate RPC functions originating from distinct ML applications. 
This capability ensures that shared remote GPUs can be utilized across multiple applications and that inference results are accurately returned to their respective sources, thereby improving performance for each application. 
However, addressing performance issues arising from resource or network contention among concurrent tasks requires multiple inference scheduling strategies (see Sec.~\ref{sec:discussion}), and consequently, the integration of these strategies into \xxx{} to further enhance its performance under such conditions remains a key objective for future work.
}

\subsection{Recording Phase Design}

\subsubsection{ML Models with Fixed-Order Operators}
\label{sec:observation}

ML models can be broadly classified into static and dynamic activation types, based on whether their sequence of activated computational operators and corresponding data flow paths remain invariant or adapt to the input during a single inference pass.

Static Activation Models (SAMs) execute a predetermined, input-invariant sequence of computational operators. 
Thus, for any given input, the specific operations and their order are constant. 
\revison{
These include:
\begin{inparaenum}[i)]
\item Multilayer Perceptrons (MLPs) and Convolutional Neural Networks (CNNs)~\cite{kapao}: These utilize fixed architectures where layers consistently perform predefined mathematical operations (e.g., matrix multiplication and convolution) regardless of input values.
\item Traditional ML models (e.g., Linear Regression~\cite{huang2020interpretable}, Support Vector Machines~\cite{wang2019gmc}, and K-Nearest Neighbors~\cite{lewis2020retrieval}): These also operate statically, their core inference logic applying a fixed set of calculations or comparisons predictable before runtime.
\item Transformer encoders~\cite{devlin2018bert}, Recurrent Neural Networks (RNNs) on fixed-length sequences~\cite{bai2018empirical}, and full Transformers on fixed-length tasks~\cite{dosovitskiy2020image}: These exhibit static behavior when processing sequences padded or truncated to a uniform, predefined length, thereby fixing the number of recurrent unrollings or activated self-attention blocks.
\end{inparaenum}
}
Characterized by fixed structures and an absence of input-dependent execution paths, these models exhibit computationally regular processes. 
This predictability makes them the primary target for our \xxx{} system and the baseline methods considered in this work.

Conversely, Dynamic Activation Models (DAMs) adapt their computational path or the number of activated operators based on the specific characteristics of the input data. 
\revison{
These include:
\begin{inparaenum}[i)]
\item Transformers (e.g., GPT-like models for generation)~\cite{brown2020language}: These exhibit dynamic behavior, particularly in auto-regressive sequence generation, where the total number of decoding steps (and thus operator activations) adapts to the runtime-determined length of the generated sequence.
\item Mixture of Experts (MoE) models~\cite{shazeer2017outrageously}: These are inherently dynamic, as a gating mechanism activates an input-dependent sparse subset of ``expert'' sub-networks (operator groups) for processing.
\item RNNs on variable-length sequences~\cite{sutskever2014sequence}: These operate dynamically (without padding), as the number of recurrent cell activations directly matches the input's sequence length, making the operator count input-dependent.
\item Decision Trees and their ensembles~\cite{chen2016xgboost}: These define dynamic inference paths, where the sequence of activated comparison operators from root to leaf is directly dictated by input features, creating input-specific computational routes.
\item Graph Neural Networks (GNNs)~\cite{jiang2019semi}: These exhibit dynamic behavior, particularly when operating on graphs of varying sizes/structures or using input-dependent neighborhood sampling. Consequently, key operational aspects adapt dynamically to each input graph or query (e.g., the number of processed nodes, the effective layer depth, and the scope of neighbor aggregation including its associated operators).
\end{inparaenum}
}
While this dynamic nature enhances model adaptability and expressiveness, it also complicates runtime profiling (e.g., execution time and I/O data sizes), leading to fewer optimization systems designed specifically for them.

\revison{
To conclude, SAMs (e.g., MLPs and CNNs for computer vision~\cite{kapao}) are widely adopted for mobile applications. 
These applications necessitate high-performance inference (low-latency and energy-efficient) but typically operate on resource-constrained mobile devices, making offloading-based inference over MEC networks essential for achieving such performance.
In contrast, DAMs, which demand significant computing resources and less stringent real-time inference requirements~\cite{brown2020language, shazeer2017outrageously}, are suited for data center deployment.}

Our system, \xxx, employs a record/replay mechanism optimized for the consistent operator sequences inherent in SAMs. 
When \xxx{} encounters a DAM where the operator sequence changes, its replayer component detects this inconsistency. 
In such instances, \xxx{} temporarily disables its specific optimizations, reverts to a standard transparent offloading workflow, and attempts to re-establish a stable execution sequence. 
Given the predominance of SAMs in mobile applications, these fallback events are expected to be infrequent, thereby preserving \xxx{}'s performance advantages. 
\revison{
Optimizing inference for DAMs with their varying operator sequences remains an open challenge for offloading systems in general. 
While some research explores solutions, such as predicting future layer activations~\cite{tarnawski2020efficient}, these advanced techniques are beyond the scope of this paper, which concentrates on SAM optimization.
}

\subsubsection{\revison{Operator Sequence Search}}
\label{sec:oss}
\revison{
A core design decision in \xxx{} is to achieve full transparency by operating without any hints from the application or the underlying ML framework. 
One might consider a simpler approach that relies on explicit markers inserted into a framework's source code (e.g., in PyTorch) to identify inference boundaries. 
However, such a method perpetuates the same platform compatibility issues that already plague device-only deployment across diverse frameworks and hardware. 
It fails to solve the fundamental deployment problem for the heterogeneous MEC ecosystem. 
Consequently, our system must autonomously discover the correct sequence of operators corresponding to an inference. 
This hint-free approach, while technically more demanding, is a prerequisite for creating a truly universal and easy-to-deploy offloading solution.
}

\begin{figure*}[htbp]
\setlength\abovecaptionskip{6pt}
\setlength\belowcaptionskip{2pt}
\centering
\includegraphics[width=0.85\linewidth]{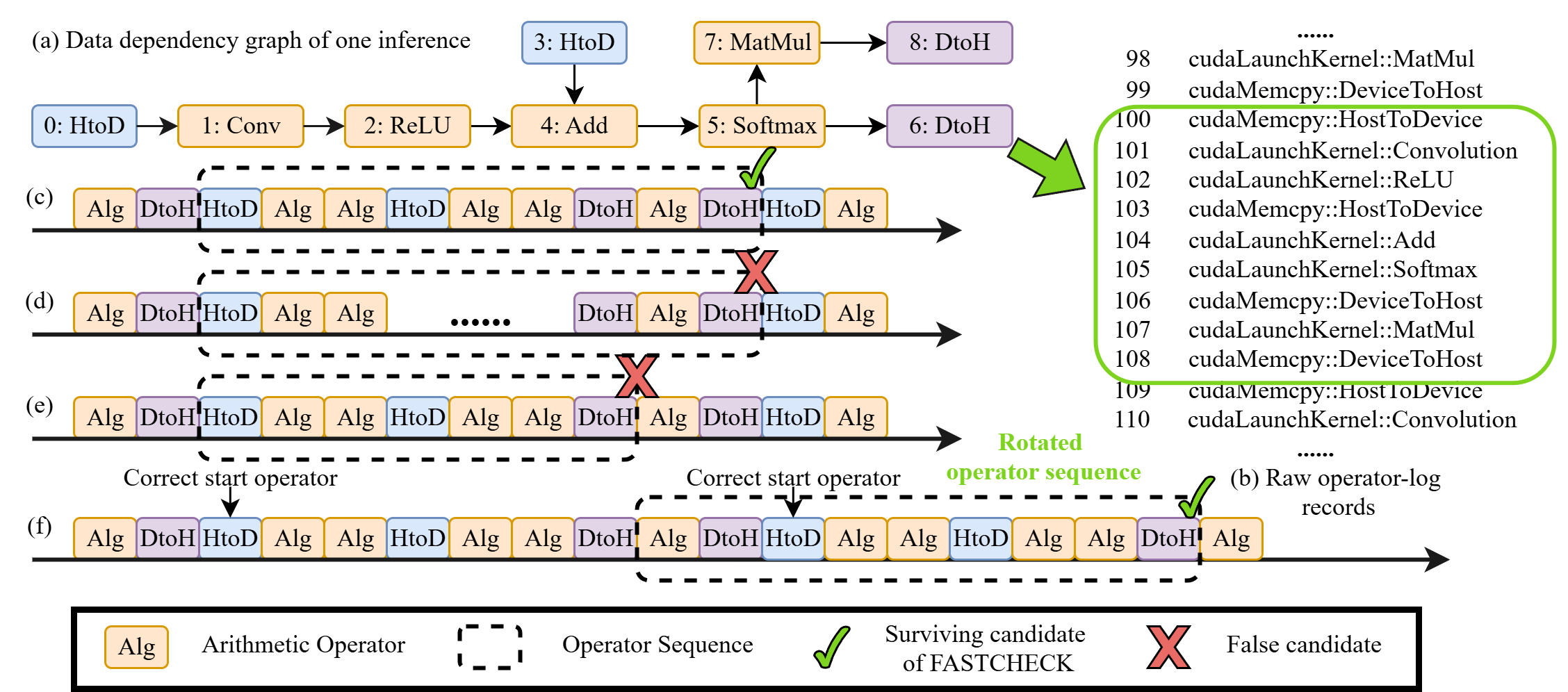}
\caption{ \small \revison{\textbf{Illustration of Operator Sequence Search.}}}
\label{fig:oss}
\end{figure*}

\revison{
The performance of \xxx{} hinges on its ability to accurately identify the correct inference operator sequence, since even a single missing or extra operator disrupts the end‐to‐end data flow and produces incorrect results. 
To achieve this under realistic conditions, Operator Sequence Search must meet three fundamental challenges:
\begin{inparaenum}[i)] 
\item  Transparency: \xxx{} cannot rely on any high-level framework hooks or metadata and must infer the sequence solely from raw operator-log records. 
\item  Initialization variability: real ML models typically emit extra or altered operators during model loading and the very first inference; these one-time initialization artifacts must not be mistaken for the regular inference pattern. 
\item Large trace size: real-world operator traces can contain tens or even hundreds of thousands of entries, so any brute-force search over all candidate subsequences would be prohibitively expensive.
\end{inparaenum} 
The pseudo code for the algorithm of operator sequence search is outlined in \textbf{Alg.~\ref{alg:oss}}, and Fig.~\ref{fig:oss} gives an example of how these challenges are addressed in our algorithm.
}

\vspace{10pt}
\begin{algorithm}[htbp]
\small
\color{black}
\caption{\small \revison{\textbf{Operator\_Sequence\_Search}}}
\label{alg:oss}
\begin{algorithmic}[1]
\Statex \textbf{Input:} \texttt{Logs}: list of OperatorInfo entries from the first $N$ inferences; $R$: minimum number of repeats for check 
\Statex \textbf{Output:} inference operator sequence \texttt{IOS} or \texttt{NULL}

\Function{OperatorSequenceSearch}{Logs, $R$}
  \State $S \gets \{v.\mathit{index}\mid v.\mathit{func}=\text{"HtoD"}, \,v \in \texttt{Logs}\,\}$ 
  \State $T \gets \{v.\mathit{index}\mid v.\mathit{func}=\text{"DtoH"}, \,v \in \texttt{Logs}\,\}$
  \If{$S = \emptyset$ or $T = \emptyset$} \Return NULL \EndIf
  \State $\text{Tags} \gets \{\,\text{Tag}(v)\mid v\in \texttt{Logs}\,\}$
  \State $end \gets \text{MAX}(T)$
  \State $starts \gets S \cup \{i+1\mid i \in T\,\}$
  \For{each $j \in starts$ \textbf{and} $j \leq end$}
    \State $\ell \gets end - j$
    \If{\Call{FastCheck}{$\text{Tags},j,\ell,R$}}
    \For{each $k \in S$ \textbf{and} $j-\ell \leq k \leq j$}
    
    \If{\Call{FullCheck}{$\texttt{Logs},k,\ell,R,T$}}
    \State $ \texttt{IOS} \gets \texttt{Logs} [\,k\;\dots\;k+\ell-1\,]$
    \State \Return $\texttt{IOS}$
    \EndIf
    \EndFor
    \EndIf
  \EndFor

  \State \Return NULL
\EndFunction
\end{algorithmic}
\end{algorithm}
\vspace{10pt}

\vspace{10pt}
\begin{algorithm}[htbp]
\small
\color{black}
\caption{\small \revison{\textbf{FastCheck\_and\_FullCheck}}}
\label{alg:fast-full}
\begin{algorithmic}[1]
\Function{FastCheck}{Tags,start, length, $R$}
  \State count $\gets 0$, pos $\gets$ start
  \While{pos $\geq 0$ \textbf{and} 
          Tags[start:start+length] = Tags[pos:pos+length]}
    \State count $\gets$ count + 1
    \State pos $\gets$ pos - length
  \EndWhile
  \State \Return count $\ge R$
\EndFunction

\Function{FullCheck}{Logs, start, length, $R$,$T$}
  \State end $\gets$ start+length-1
  \If{ end $\notin T$ \textbf{or} \Call{CheckDataDependency}{$Logs,start,length$} $= \texttt{False}$}
  \State \Return $\texttt{False}$
  \EndIf
  \State count $\gets 0$, pos $\gets$ start
  \While{pos $\geq 0$}
    \If{for all $0 \le t < \text{length}$,\ 
         \texttt{Logs}[start+t] $\equiv$ \texttt{Logs}[pos+t]}
      \State count $\gets$ count + 1
      \State pos $\gets$ pos - length
    \Else
      \State \textbf{break}
    \EndIf
  \EndWhile
  \State \Return count $\ge R$
\EndFunction

\end{algorithmic}
\end{algorithm}
\vspace{20pt}

\revison{
To address these challenges, our algorithm exploits three key observations. 
\textbf{\textcircled{1}} A static ML model executes the same operator sequence for each inference, yielding a highly regular pattern in the logs. 
\textbf{\textcircled{2}} The real-time requirements of MEC enforces processing inference immediately, which begins with a host‐to‐device memory copy of the raw input (``cudaMemcpyHtoD'', short for ``HtoD'') and ends with a device-to-host memory copy of the result (``cudaMemcpyDtoH'', short for ``DtoH''). 
By treating these copies as special memory transfer operations and grouping any following synchronization calls with them (e.g., ``cudaStreamSynchronize''), we obtain reliable boundary markers.
\textbf{\textcircled{3}} The correct inference sequence must satisfy all data-dependency constraints: every operator’s inputs must originate either from the raw input, from a prior operator’s output (share the same memory address), or from the model’s parameters.}

\revison{
Unfortunately, relying solely on the maximum-repeated-substring algorithm~\cite{charalampopoulos2021faster} based on observation \textbf{\textcircled{1}} fails to segment the operator sequence correctly, because when the sequence repeats continuously in the log, the algorithm merges multiple consecutive iterations into a single maximal substring (Fig.~\ref{fig:oss}d).
Likewise, using only the memory-copy markers of observation \textbf{\textcircled{2}} falls short when multiple ``HtoD'' or ``DtoH'' events occur within a single inference, leaving ambiguous which copy corresponds to the true beginning or end (Fig.~\ref{fig:oss}e).}

\revison{
Based on these observations, we introduce a robust two-stage matching strategy. 
This strategy is engineered to systematically reduce the search complexity, first by rapidly identifying plausible candidates and subsequently by performing detailed verification on this narrowed selection, thereby accelerating the overall identification process.}

\revison{
In the first stage, \texttt{FASTCHECK}  scans the log to identify candidate start and end operators based on memory‐copy boundaries of observation \textbf{\textcircled{2}} (\textit{Lines 2 and 3 in Alg.~\ref{alg:oss}}) and then locates possible operator sequences by detecting repeated sequences based on observation \textbf{\textcircled{1}} (\textit{Line 3 in Alg.~\ref{alg:fast-full}}). 
Each candidate ends at the current last end operator (\textit{Line 7 in Alg.~\ref{alg:oss}}); if that operator truly marks the inference end, the candidate begins at the matching start operator (Fig.~\ref{fig:oss}c), whereas if it lies within a rotated sequence, the candidate instead begins at the next occurrence of an end operator (Fig.~\ref{fig:oss}f) (\textit{Line 8 in Alg.~\ref{alg:oss}}). 
\texttt{FASTCHECK} then linearizes the log into a compact string of operator categories (``HtoD'', ``DtoH'', arithmetic, etc.) that it can count in linear time how often the candidate substring reappears in the latter portion of the log (\textit{Line 6 in Alg.~\ref{alg:oss}}) and thereby confirm sustained repetition despite initialization variability (\textit{Line 3 in Alg.~\ref{alg:fast-full}}). 
By focusing solely on category tags and recurrence, \texttt{FASTCHECK} prunes the vast majority of false candidates caused by one‐time initialization artifacts or spurious memory‐copy boundaries.
}

\revison{
Once \texttt{FASTCHECK} has identified the surviving candidates, operator sequence search advances to the second stage, \texttt{FULLCHECK}, which verifies each candidate exhaustively (\textbf{Alg.~\ref{alg:fast-full}}). 
Because a \texttt{FASTCHECK} candidate may be a cyclic rotation of the true sequence, \texttt{FullCheck} first realigns it to the correct start and end markers using the host‐to‐device and device‐to‐host boundaries from observation \textbf{\textcircled{2}} (Fig.~\ref{fig:oss}f, \textit{Line 12 in Alg.~\ref{alg:oss}}).
It then checks that every operator in the sequence satisfies the required data dependencies according to observation \textbf{\textcircled{3}} (\textit{Line 11 in Alg.~\ref{alg:fast-full}}) that inputs must come from raw input, a prior operator’s output at the same address, or model parameters. 
Finally, it conducts a full one-to-one, record-level comparison across the entire log to confirm that the sequence repeats exactly as predicted by observation one (\textit{Line 16 in Alg.~\ref{alg:fast-full}}). 
Although this final validation is the most time-consuming step, it is applied only to a small set of high-confidence candidates.
}

\revison{
In this way, the operator sequence search algorithm addresses the three challenges, and locates the correct operator sequence efficiently by keeping transparency through pure log analysis, discounting one‑time initialization variability via repeated‑sequence detection, and taming large traces with a two‑stage \texttt{FastCheck}+\texttt{FullCheck} strategy. 
This design enables \xxx{} to extract the precise inference operator sequence without any framework support, even in demanding real-world deployments.
}



\subsection{Replaying Phase Design}
\subsubsection{Workflow of Replaying Phase}
\label{sec:replay}
\begin{figure}[htbp]
\setlength\abovecaptionskip{6pt}
\setlength\belowcaptionskip{5pt}
\centering
\includegraphics[width=0.85\linewidth]{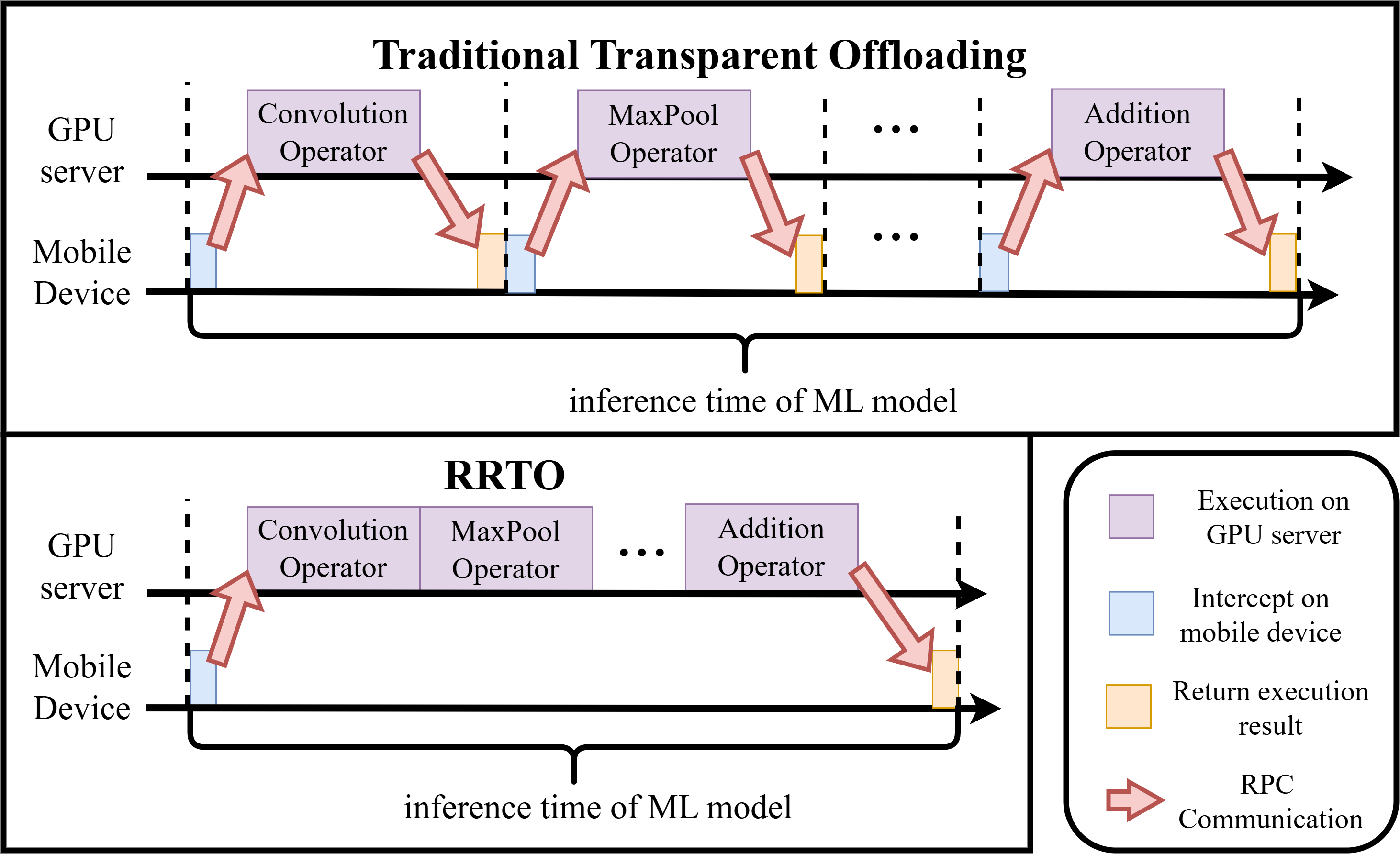}
\caption{ \small \revison{\textbf{Workflow of \xxx{} during the replaying phase.}}}
\label{fig:workflow}
\end{figure}

In this section, we present the workflow of \xxx{} during the replaying phase and explain how it eliminates the per-operator RPC communication required by existing transparent offloading methods via its record/replay mechanism. 
Fig.~\ref{fig:workflow} illustrates this workflow and compares it with traditional transparent offloading systems during model inference in MEC networks. 
Traditional transparent offloading suffers from frequent operator-level RPC communication, leading to high communication overhead and degraded system performance (see Sec.~\ref{sec:evaluation}), including lower GPU utilization on servers, longer inference times, and increased energy consumption per inference.

To address the communication costs in the transparent offloading systems, \xxx{} introduces an automatic recording and replay mechanism. 
Given that ML models in mobile applications often exhibit static and predictable sequences of operator invocation during inference, \xxx{} records the operators called during the initial inferences and replays this recorded sequence for subsequent inferences. 
As illustrated in Fig.~\ref{fig:workflow}, during the replay phase, \xxx{} only offloads the start operator in the sequence via RPC as done in traditional transparent offloading systems, streamlining the start of the inference task. 
Subsequent operators are directly invoked on the GPU server side, instead of being invoked by RPCs from the mobile device side as in traditional systems. 
Consequently, \xxx{} executes all required operators within the inference operator sequence in one shot at the beginning of each inference, eliminating the need for additional RPC communications for subsequent operators and significantly reducing communication costs.

\subsubsection{Record/Replay Mechanism}
\label{sec:pseudo}

Here we describe how \xxx{} implements its record/replay mechanism. 
The transparent offloading process is outlined in two parts: the client component is detailed in \textbf{Alg.~\ref{alg:client}}, and the server component in \textbf{Alg.~\ref{alg:server}}. 

\vspace{10pt}
\begin{algorithm}[htbp]
\small
\caption{\small \textbf{\xxx\_on\_Client}}\label{alg:client}
\begin{algorithmic}[1]
\Statex \textbf{Input:}  CUDA API called by the corresponding operator $\texttt{func}$ and the required parameters $\texttt{args}$
\Statex \textbf{Output:} The execution result $\texttt{ret}$
\Statex \textbf{Parameter:} inference operator sequence $\texttt{IOS}$
\State $\texttt{IOS} \gets \emptyset$, $\texttt{Logs} \gets \emptyset$
\While {True}
\If{$\texttt{IOS}$ is NULL}
    \State // recorder
    \State \Call{SendRPCtoServer}{$\texttt{func},\texttt{args}$}
    \State $\texttt{IOS} \gets$ \Call{OperatorSequenceSearch}{$\texttt{Logs}$}
    \State $\texttt{ret} \gets $\Call{GetRPCExecutionResult}{$\texttt{}$}
    \State $\texttt{Logs} \gets \texttt{Logs} \cup \{(\texttt{func},\texttt{args},\texttt{ret})\}$
\Else
    \State // replayer on mobile device
    \If{$\texttt{func}$ is $\texttt{IOS}[0][``func"]$}
    \State \Call{StartRRTO}{$\texttt{IOS}$}
    \EndIf
    \If{$\texttt{func}$ is ``HtoD''}
    \State $\texttt{ret} \gets$ \Call{SendRPCtoServer}{$\texttt{args}$}
    \ElsIf{$\texttt{func}$ is ``DtoH''}
    \State $\texttt{ret} \gets$ \Call{WaitingForRRTO}{$\texttt{}$}
    \Else
    \State $idx \gets $ \Call{find}{$\texttt{IOS},\texttt{func}$}
    \State $\texttt{ret} \gets \texttt{IOS}[idx][``ret"]$
    \EndIf
\EndIf
\State \Call{ReturnResult}{$\texttt{ret}$}
\EndWhile
\end{algorithmic}
\end{algorithm}

\vspace{10pt}
In \textbf{Alg.~\ref{alg:client}} on the offloading client side, \xxx{} takes a CUDA kernel function called by an operator and the requisite parameters as input and  keep serving each operator. 
Initially, the algorithm checks whether the recorder has already identified the inference operator sequence to determine if the current phase should be recording or replaying.  
If the sequence has not been established, \xxx{} enters the recording phase (\textit{Lines 4 to 8}), which involve sending an RPC to the server (\textit{Line 5}), performing the operator sequence search (\textit{Line 6}), and capturing and recording the RPC execution result (\textit{Lines 7 to 8}), adhering to the execution pattern of traditional transparent offloading systems. 
To enhance system efficiency, \xxx{} overlaps the operator sequence search with the RPC execution, allowing the algorithm to complete while the client awaits the RPC result.
\revison{In this context, $func$ denotes a specific CUDA API call (e.g., ``cudaLaunchKernel'', ``cudaMemcpyHtoD'', and ``cudaMemcpyDtoH''). 
The term $args$ represents the required parameters for that call (e.g., CUDA kernel configurations and memory addresses of parameters). 
Finally, $ret$ captures the execution result, which is typically a status code like ``cudaSuccess'' from ``cudaLaunchKernel''.}

If the inference operator sequence is already identified, \xxx{} transitions to the replaying phase on the mobile device (\textit{Lines 10 to 20}). 
This phase starts by initiating \xxx{} for a new inference at the start operator (\textit{Line 12}), then returns the execution results of previous RPC calls for intermediate operators within the sequence (\revison{mainly `cudaSuccess' from ``cudaLaunchKernel''}, \textit{Lines 19 to 20}), synchronizes inference input at ``cudaMemcpyHtoD'' (\textit{Line 15}) and inference output at ``cudaMemcpyDtoH'' (\textit{Line 17}).
Finally, it returns the execution result of each function to the upper-layer applications (\textit{Line 23}).

\vspace{10pt}
\begin{algorithm}[htbp]
\caption{ \small \textbf{\xxx\_on\_Server}}\label{alg:server}
\begin{algorithmic}[1]
\small
\Statex \textbf{Input:} client task $\texttt{task}$
\Statex \textbf{Parameter:} inference operator sequence $\texttt{IOS}$, the execution result $\texttt{ret}$
\State  $\texttt{IOS} \gets \emptyset$
\While {True}
\If{$\texttt{task}$ is \Call{SendRPCtoServer}{}}
    \State $\texttt{func},\texttt{args} \gets$ \Call{GetClientInput}{$\texttt{}$}
    \State $\texttt{ret} \gets$ \Call{CUDARuntimeLibrary}{$\texttt{func},\texttt{args}$}
\ElsIf{$\texttt{task}$ is  \Call{StartRRTO}{}}
    \State //  replayer on server
    \State $\texttt{IOS} \gets$ \Call{GetClientInput}{$\texttt{}$}
    \ForAll {$\texttt{Op} \in \texttt{IOS}$ }
    \State $\texttt{func} \gets \texttt{Op}[``func'']$, $\texttt{args} \gets \texttt{Op}[``args'']$
    \If{$\texttt{func}$ is ``HtoD''}
     \State $\texttt{input} \gets$ \Call{GetClientInput}{$\texttt{}$}
    \State $\texttt{args} \gets$ \Call{RRTOFixArgs}{$\texttt{args},\texttt{input}$}
    \EndIf
    \State $\texttt{ret} \gets$ \Call{CUDARuntimeLibrary}{$\texttt{func},\texttt{args}$}
    \If{$\texttt{func}$ is ``DtoH''}
    \State \Call{SendExecutionResultBack}{$\texttt{ret}$}
    \EndIf
    \EndFor
\EndIf
\EndWhile
\end{algorithmic}
\end{algorithm}
\vspace{10pt}

In \textbf{Alg.~\ref{alg:server}}, the \xxx{} offloading server continuously awaits tasks from the client, aligning its operational phase with that of the client to ensure synchronized processing. 
During the recording phase, the server processes RPC requests similarly to traditional transparent offloading systems (\textit{Lines 3 to 5}).
As the client on the mobile device initiates its replaying phase, the offloading server simultaneously begins its own (\textit{Lines 7 to 17}), effectively replaying the execution of the recorded inference operator sequence. 
During this phase, \xxx{} configures each operator’s parameters (\textit{Line 13}), typically supplying the current inference input data or its memory addresses, to ensure accurate and efficient execution.

\section{Implementation}
\label{sec:implement}
In this section, we first elaborate on the implementation of \xxx, and then introduce the experiment setup.
\subsection{Implementing \xxx}
\myparagraph{Software}
We implemented \xxx{} within Cricket's codebase~\cite{cricket}, a transparent offloading system that provides a virtualization layer for CUDA applications, enabling remote execution without the need for source code modifications and recompilation of applications. 
\xxx{} employs the same RPC library for communication operations as Cricket: Libtirpc \cite{libtirpc}, a transport-independent RPC library for Linux.
We integrated \xxx's recorder and replayer into the corresponding RPC functions in Cricket, allowing for seamless integration and efficient operation of the record/replay mechanism.

\begin{figure}[!t]
\setlength\abovecaptionskip{6pt}
\setlength\belowcaptionskip{2pt}
\centering
\includegraphics[width=0.85\linewidth]{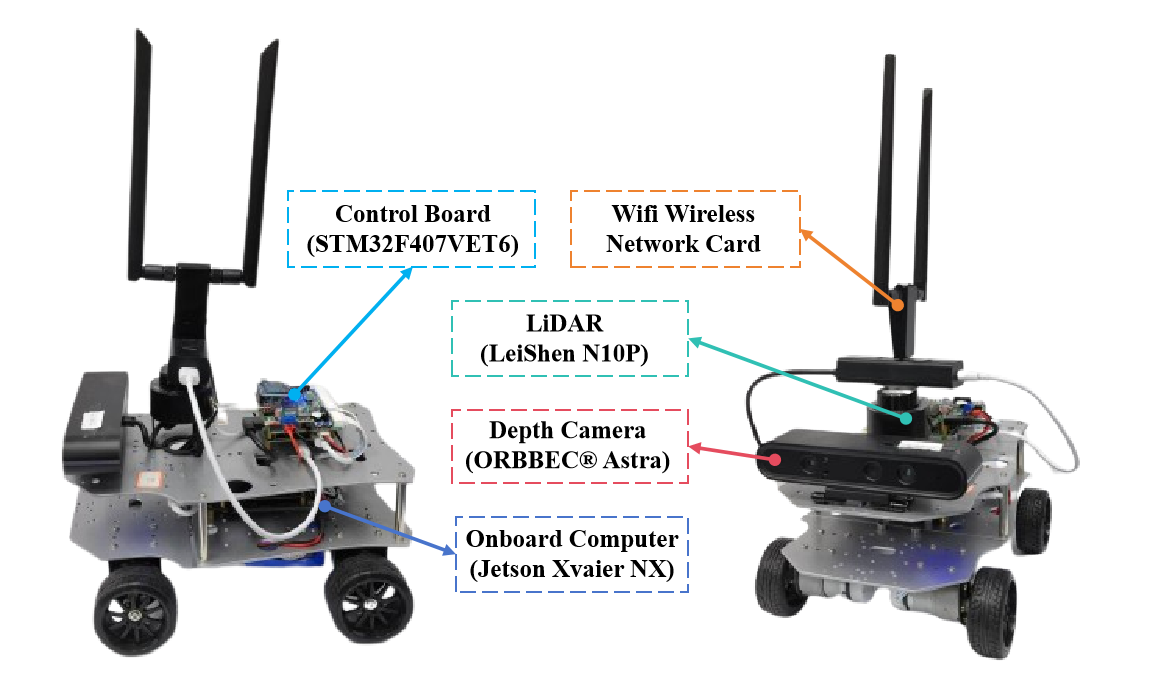}
\caption{ \small \textbf{The detailed composition of the robot platform.}}
\label{fig:robot}
\end{figure}

\myparagraph{Hardware} 
The evaluation was conducted on a customized four-wheeled robot (Fig.~\ref{fig:robot}), equipped with a Jetson Xavier NX~\cite{jetsonnx} 8G onboard computer that is capable of CUDA-acclerated ML model inference. 
The system runs Ubuntu 20.04 with ROS Noetic and uses a dual-band USB network adapter (MediaTek MT76x2U) for wireless communication. 
Detailed hardware and sensor configurations are shown in Fig.~\ref{fig:robot}. 

The GPU server is a PC equipped with an Intel(R) i5 12400f CPU @ 4.40 GHz and an NVIDIA GeForce GTX 2080 Ti 11 GB GPU, connected to our robot via Wi-Fi 6 over a 160 MHz channel at 5 GHz frequency.

\begin{table}[!t]
\setlength\abovecaptionskip{6pt}
\setlength\belowcaptionskip{2pt}
\centering
\begin{tabular}{|c|c|c|c|}
\hline
        & inference & communication & standby \\ \hline
 Energy (Watt) &     13.35        &       4.25        &    4.04   \\ \hline
\end{tabular}
\caption{ \small \textbf{Power draw (Watt) of our robot in different states.}}
\label{tab:energydefault}
\end{table}

Tab.~\ref{tab:energydefault} summarizes the robots' on-board energy consumption (excluding motor power) in different states: inference (full GPU utilization, including CPU/GPU power), communication (communication with the GPU server, including wireless network card energy consumption), standby (no tasks to execute).
Each Jetson Xavier NX is powered by a 21.6 Wh battery, sustaining up to 1.6 hours of continuous model inference. 
\revison{We continuously log the robot's instantaneous on-board power draw (in Watts) at 1-second intervals, utilizing the back-end power consumption and performance monitoring methodology from~\cite{jetsonnx}.
Subsequently, the energy consumption per inference (in Joules) is precisely calculated by integrating this power draw profile over the exact duration of each inference, determined by its start and end timestamps.}

\subsection{Experiment Setup}

\begin{figure}[!t]
\setlength\abovecaptionskip{6pt}
\setlength\belowcaptionskip{-2pt}
    \centering
    \includegraphics[width=0.85\linewidth]{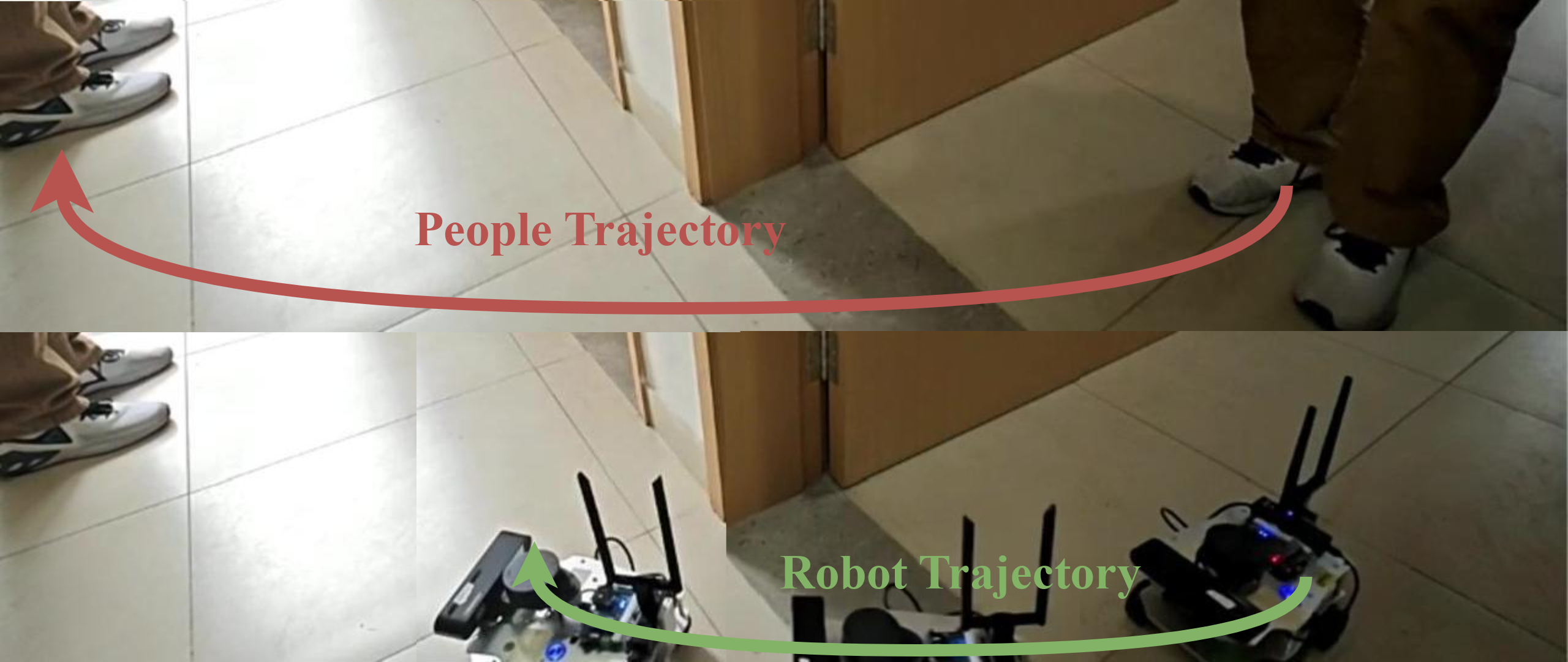}
    \caption{\small \textbf{Kapao~\cite{kapao}}, a real-time people-tracking application on our four-wheeled robot with a CNN-based human keypoint detection model.}
    \label{fig:real-world}
\end{figure}

\begin{figure}[!t]
\setlength\abovecaptionskip{6pt}
\setlength\belowcaptionskip{2pt}
\centering
\includegraphics[width=0.85\linewidth]{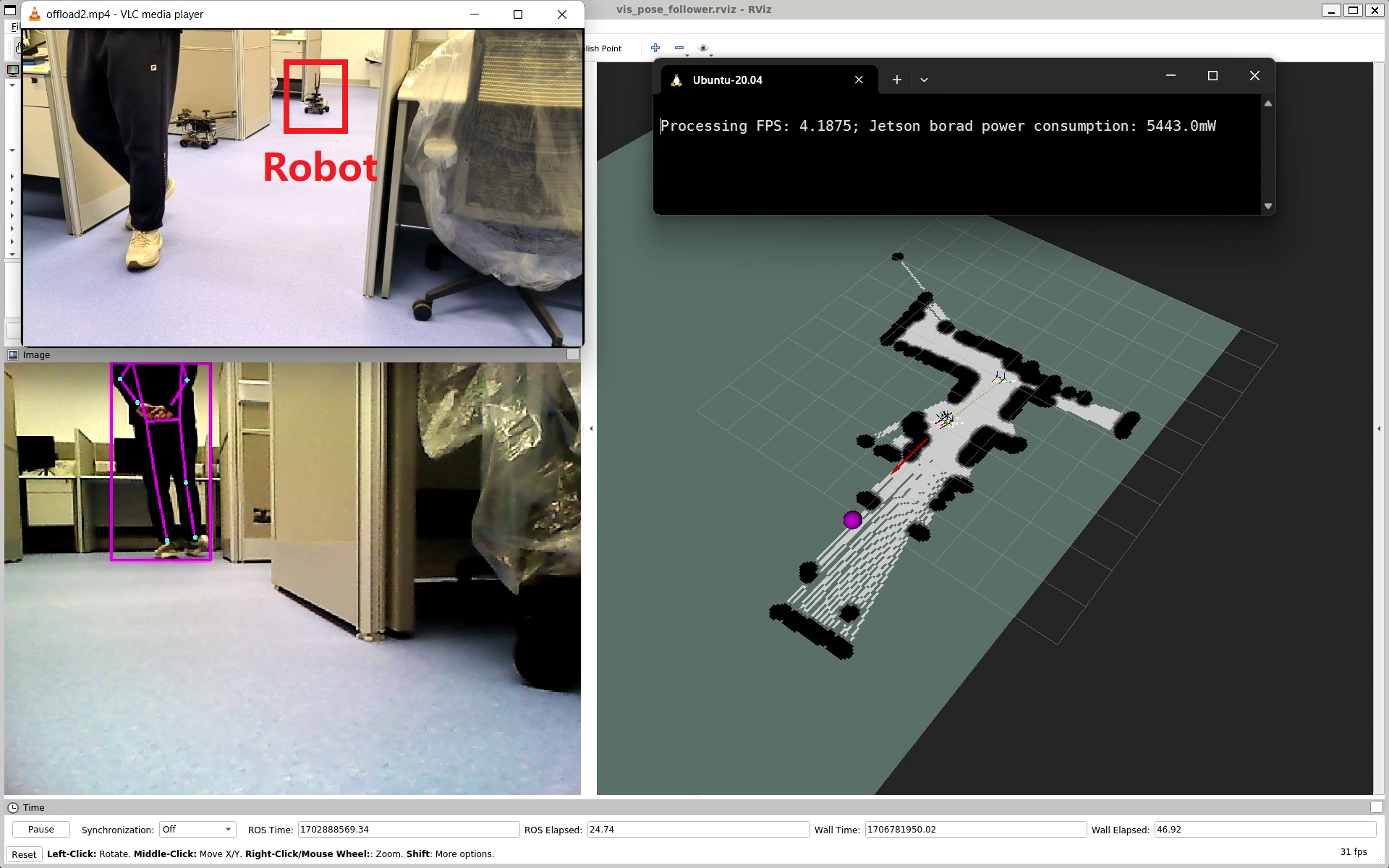}
\caption{ \small A screenshot of our real-world experiment. The upper right corner displays real-time FPS and on-board energy consumption, the lower right corner shows the map created by the robot using its LiDAR, the lower left corner features the real-time view from the robot's camera, and the upper left corner provides a third-angle observation of the entire experimental process.}
\label{fig:workload}
\end{figure}

\myparagraph{Task}
We evaluated a real-time people-tracking robotic application on our robot as depicted in Fig.~\ref{fig:real-world} and Fig.~\ref{fig:workload}.
To demonstrate the generalization ability of \xxx{}, we also evaluated several representative models from three categories of real-world mobile applications with their implementations available in Torchvision~\cite{torchvision}: 
\begin{inparaenum}[i)] 
\item  ResNet~\cite{he2015deepresiduallearningimage} and ConvNext~\cite{liu2022convnet2020s} for 
object classification; 
\item FCN~\cite{long2015fullyconvolutionalnetworkssemantic} and DeepLabv3~\cite{chen2017rethinkingatrousconvolutionsemantic} for semantic segmentation;
\item FasterRCNN~\cite{ren2016fasterrcnnrealtimeobject} and RetainNet~\cite{lin2018focallossdenseobject} for object detection.
\end{inparaenum}

\myparagraph{Emulation Environments}
We evaluated two real-world environments: indoors (robots move in our laboratory with desks and separators interfering with wireless signals) and outdoors (robots move in our campus garden with trees and bushes interfering with wireless signals, resulting in lower bandwidth). 
The corresponding bandwidths between the robot and the GPU server in indoors and outdoors scenarios are shown in Fig.~\ref{fig:bandwidth}.

\myparagraph{Baselines}
To comprehensively evaluate the performance of \xxx, we conducted comparative experiments against several baseline approaches:
\revison{
\begin{itemize}
    \item \textbf{Device-only inference (``Device-only'')}: A conventional setup where the entire model is deployed and executed on the robot.  
    \item \textbf{Native non-transparent offloading (``NNTO'')}: A non-transparent offloading method that deploys the entire model on a GPU server by modifying the source code. 
    We evaluate NNTO independently to demonstrate the benefits of offloading in inference latency and energy efficiency, and we use it as the theoretical upper bound for offloading approaches because it incurs minimal system transmission overhead by transmitting only inference inputs and outputs. 
    This comparison underscores the communication time savings and performance gains achieved by \xxx.
    \item \textbf{Cricket}~\cite{cricket}: A state-of-the-art transparent offloading system designed for remote GPU usage.
\end{itemize}  
}


\revison{
While advanced scheduling optimizations (e.g., layer partitioning~\cite{kang2017neurosurgeon} and multiple inference scheduling~\cite{lin2019computation, fang2017qos, zhao2019novel}) are adaptable to \xxx{}, their performance benefits are orthogonal to our core contributions. Consequently, to isolate the impact of our innovations and ensure a fair comparison against existing non-transparent offloading (NNTO) methods, we have excluded these optimizations from our current implementation and evaluation. Their integration into \xxx{} remains a promising direction for future work, as discussed in Sec.~\ref{sec:discussion}.
}

\section{Evaluation}
\label{sec:evaluation}
In this section, we evaluate the performance of \xxx{} from four aspects:
\revison{
\begin{inparaenum}[i)] 
\item  comparison with baseline systems on inference latency and energy consumption in real-world mobile application; 
\item sensitivity of \xxx{} in various MEC scenarios;
\item analysis of \xxx{}’s record/replay mechanism;
\item performance evaluation of \xxx{} on large-scale models common to fundamental mobile applications.
\end{inparaenum} 
}

\subsection{Superiority of \xxx}

\begin{figure}[htbp]
\setlength\abovecaptionskip{6pt}
\setlength\belowcaptionskip{2pt}
\centering
\includegraphics[width=0.98\linewidth]{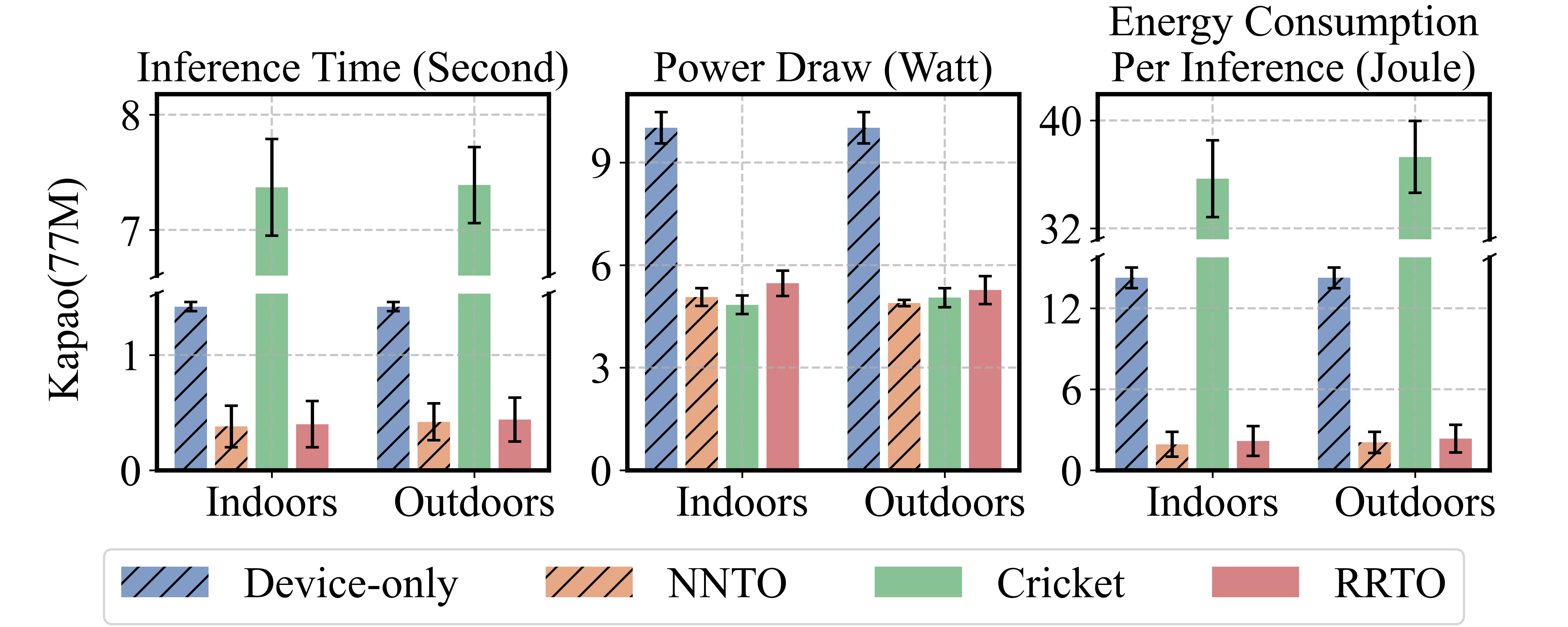}
\caption{ \small \revison{\textbf{Performance of Kapao in different environments with various systems.}}}
\label{fig:kapao}
\end{figure}

Our evaluation results of our robotic application, KAPAO, as presented in Fig.~\ref{fig:kapao}, demonstrate that \xxx{} achieved performance comparable to non-transparent offloading system (NNTO) in both inference time and energy consumption.

In terms of inference time, \xxx{} reduced inference time by an average of 72\% compared to local computation and 95\% compared to Cricket in the indoors scenario; the reductions in the outdoors scenario were 69\% and 94\%, respectively.
Device-only, which processes the entire model on the robot without any data transmission to a GPU server, exhibited consistent performance in both indoor and outdoor scenarios. 
In contrast, the substantial communication costs associated with Cricket's frequent RPCs from its transparent offloading mechanism considerably slowed its inference times, a point that will be elaborated on in Sec.~\ref{sec:micro}.
By implementing its innovative record/replay mechanism, \xxx{} effectively minimized these extensive communication costs, achieving inference times comparable to those of NNTO, with nearly identical communication expenses.

In terms of energy consumption, \xxx{} achieved substantial reductions, decreasing energy usage per inference by an average of 85\% compared to local computation and 94\% compared to Cricket in indoors scenarios; outdoors reductions were 84\% and 93\%, respectively. 
Due to the intensive computational demands of the model, device-only inference incurred high power draw, whereas other systems that offload computations to a GPU server exhibited reduced energy usage. 
Although \xxx{} only reduced power draw by 45\% compared to local computation and even showed a 13\% increase in power draw compared to Cricket, its shorter inference times resulted in significantly lower energy consumption per inference. 
It is important to note that the average power draw values shown in Fig.~\ref{fig:kapao} do not correspond to those in Tab.~\ref{tab:energydefault}. 
This discrepancy arises because our application does not fully utilize the GPU capabilities of the Jetson Xavier NX, resulting in lower average energy consumption during local computation than during the inference stage. 
Furthermore, additional CPU computing tasks (e.g., robot control) cause the average energy consumption of all offloading systems to increase above levels observed during communication and standby phases.

The prolonged inference times observed in outdoor scenarios for all offloading systems can be attributed to the lower bandwidth available outdoors (see Sec.~\ref{sec:background-transparent}), which results in extended transmission times compared to indoor scenarios. 
Analyzing performance in both indoors and outdoors settings, we find that \xxx{} is robust across various MEC scenarios. 
This robustness stems from \xxx's ability to eliminate the frequent transmission requirements of RPCs in Cricket, thereby reducing communication overhead to levels comparable to NNTO.
\revison{Moreover, when GPU servers reside in commercial cloud environments, network congestion and routing inefficiencies further restrict available bandwidth~\cite{noormohammadpour2017datacenter} and drive up transmission costs, making \xxx{} even more advantageous than Cricket.}


\subsection{Micro-Event Analysis}
\label{sec:micro}


\begin{table*}[!t]
\setlength\abovecaptionskip{6pt}
\setlength\belowcaptionskip{-10pt}
    \centering
    \begin{tabular}{|c|c|c|c|}
    \hline
   CUDA Runtime API & \makecell{Composition during\\loading model} & \makecell{Composition during\\initializing inference} & \makecell{Composition during\\the following inference loop} \\ \hline
cudaGetDevice &46858 (82.32\%) &4789 (80.12\%) & 4735 (80.32\%) \\ \hline
cudaGetLastError &4244 (7.46\%) & 616 (10.31\%)& 607 (10.30\%) \\ \hline
cudaLaunchKernel & 2752 (4.83\%) & 523 (8.75\%) & 522 (8.85\%) \\ \hline
cudaMalloc &65 (0.11\%) & 4 (0.07\%) & 0 (0.00\%) \\ \hline
cudaStreamIsCapturing &68 (0.12\%) & 4 (0.07\%) & 0 (0.00\%) \\ \hline
cudaStreamSynchronize &1118 (1.96\%) & 16 (0.27\%) & 11 (0.19\%) \\ \hline
cudaMemcpyHtoD & 1117 (1.96\%) & 7 (0.12\%) & 3 (0.05\%) \\ \hline
cudaMemcpyDtoH & 1 (0.002\%) & 9 (0.15\%) & 8 (0.14\%) \\ \hline
cudaMemcpyDtoD & 701 (1.23\%) & 9 (0.15\%) & 9 (0.15\%) \\ \hline
\end{tabular}
\caption{ \small \textbf{Composition of RPC function calls during different stages of KAPAO inference.}}
\label{tab:Micro_RPC}
\end{table*}


To gain a deeper insight into the performance improvements facilitated by \xxx, we analyzed the RPC function calls made by Cricket during various stages of KAPAO inference, illustrating the characteristics of traditional transparent offloading mechanisms. 
A detailed breakdown of these calls is provided in Tab.~\ref{tab:Micro_RPC}.

Comparing the different stages of function calls in Tab.~\ref{tab:Micro_RPC}, we can see that KAPAO undergoes an initialization stage of inference different from subsequent inference loops.
This is because the working process of KAPAO~\cite{kapao} follows the default detection model in Yolo v5~\cite{glenn_jocher_2022_7347926}: the inference pipeline is first initialized by generating a mesh grid of a certain size that fits the input image size, which serves as the storage of intermediates; then in the following loop iterations through the inference pipeline the mesh grid is reused and the operator call sequence is fixed.
\xxx{} records all involved operators during the first few inferences, not just the initial process, and ignores the different operator sequences from the initializing inference until the correct operator sequence is found.

In the loop inference detailed in Tab.~\ref{tab:Micro_RPC}, we observed that a significant portion, specifically 90.62\%, of RPC function calls consisted of ``cudaGetDevice'' and ``cudaGetLastError''. 
These calls, generated by PyTorch~\cite{pytorch} due to our application's reliance on this framework, are crucial for determining the data's location, facilitating computations across multiple GPUs and parallel tasks. 

\begin{figure}[htbp]
\setlength\abovecaptionskip{6pt}
\setlength\belowcaptionskip{2pt}
\centering
\includegraphics[width=0.98\linewidth]{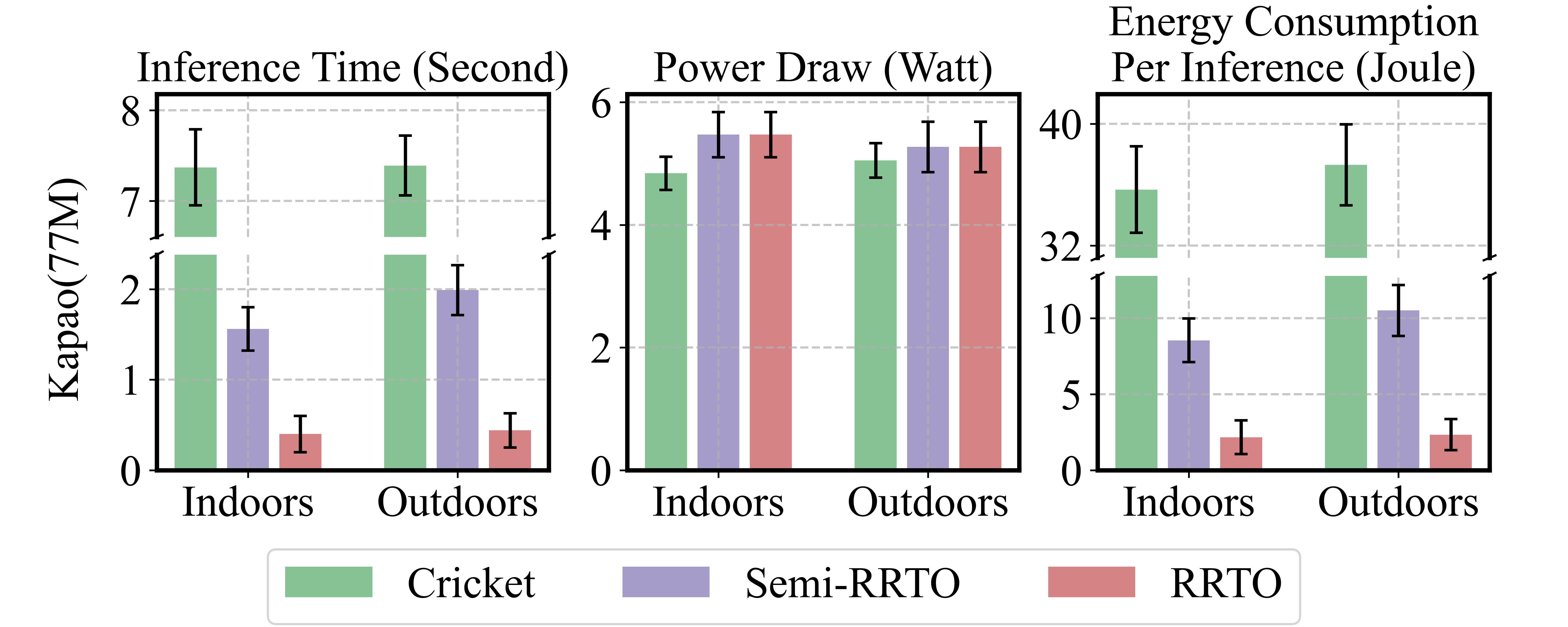}
\caption{ \small \revison{\textbf{Semi-RRTO}: only applying Caching~\cite{singhvi2021cliquemap} specifically to the RPCs of ``cudaGetDevice'' and ``cudaGetLastError'' in \xxx, effectively eliminating their transmission requirements.}}
\label{fig:semi}
\end{figure}

Despite restricting PyTorch to use only a single GPU sequentially and employing Caching (referred to as ``semi-RRTO'' in Fig.~\ref{fig:semi}) to reduce the transmission of RPCs, semi-RRTO achieved inference time comparable only to local computation in our experiments and did not reach the speeds observed with NNTO. 
This is evident from the fact that ``cudaLaunchKernel'' still represents 8.85\% of total RPC function calls, which are essential for notifying the server about subsequent computing tasks like additional convolution or maxpool operations.
While traditional RPC optimization methods wait for ``cudaLaunchKernel'' RPCs from the client to direct the server's subsequent computing tasks (operators), \xxx{} recorded these ``cudaLaunchKernel'' function calls and directly executed the subsequent computing tasks on the server, thereby eliminating the need for ongoing communication with the client.

\begin{figure*}[!t]
\setlength\abovecaptionskip{3pt}
\setlength\belowcaptionskip{-10pt}
\centering
\includegraphics[width=0.98\linewidth]{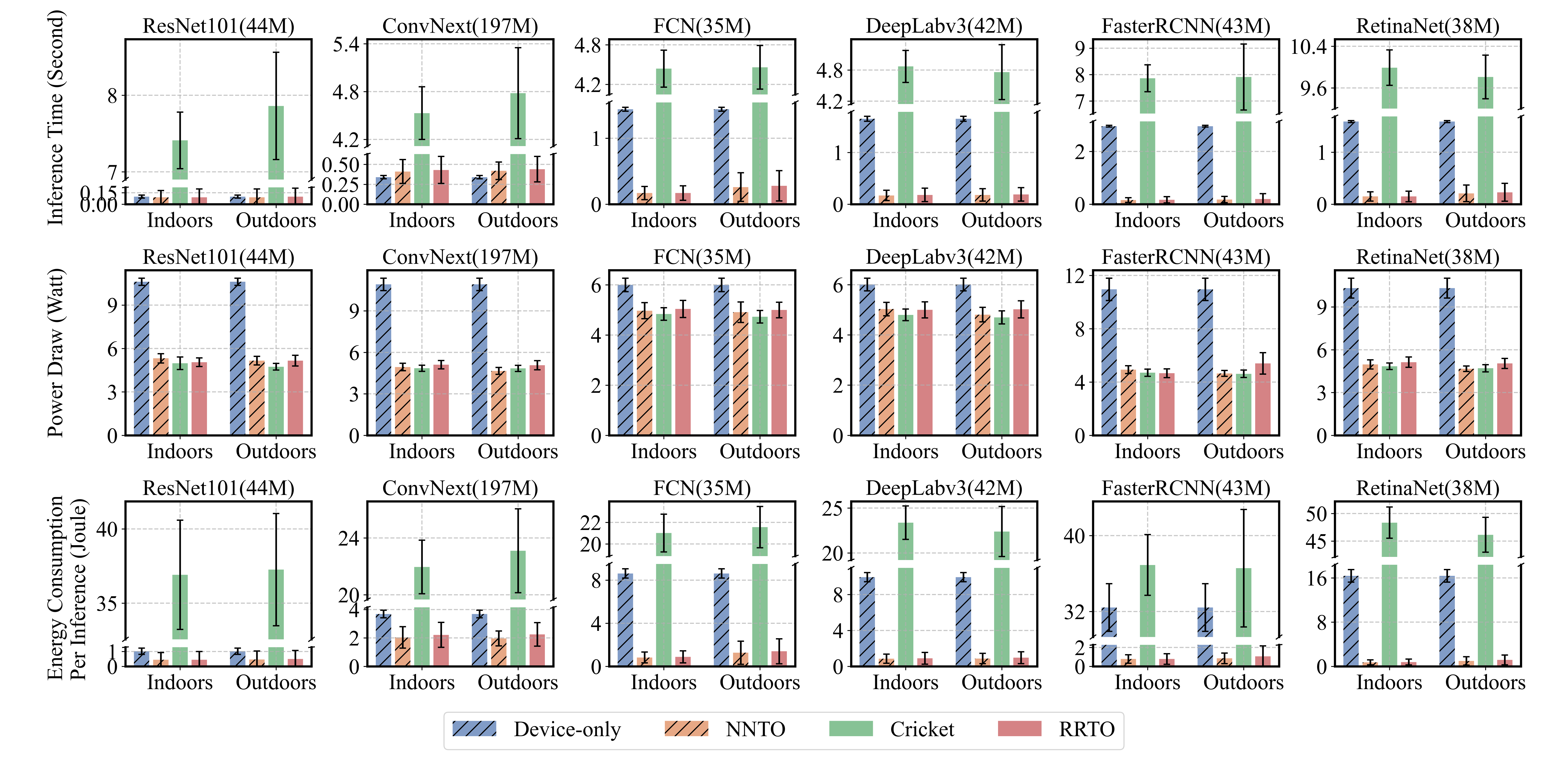}
\caption{ \small \revison{\textbf{Performance of Torchvision models in different environments with various systems.}}}
\label{fig:torchvision}
\end{figure*}

Regarding the remaining RPC functions, namely ``cudaMalloc'', ``cudaStreamIsCapturing'', ``cudaStreamSynchronize'', and ``cudaMemcpyDtoD'', which collectively account for 0.34\% of the total RPC calls, they primarily handle data transmission and synchronization within the GPU and can also be replayed by \xxx{} on the server. 
However, ``cudaMemcpyHtoD'' and ``cudaMemcpyDtoH'', which account for 0. 19\% of total RPC calls, are used primarily for data transmission between the CPU and GPU.
They are mainly used for the input and output of the ML model and cannot be replayed by \xxx{}.

To further illustrate the performance gains achieved by \xxx, we compared it with baseline systems in the number of RPC calls and the resulting average GPU utilization on the GPU server during the execution of~\cite{kapao}, measured using pynvml~\cite{corporation_pynvml_nodate} and presented in Tab.~\ref{tab:Micro}. 
Unlike \xxx{} and Cricket, NNTO bypassed RPC by directly synchronizing the input and output data of the ML model between the CPU and the GPU at the application layer, requiring modifications to the source code. 
Cricket, on the other hand, experienced higher communication costs, which contribute to lower GPU utilization on the GPU server. 
Although \xxx{} also managed ``cudaMemcpyDtoH'' and ``cudaMemcpyHtoD'' like Cricket, resulting in 11 RPCs per inference, the performance improvements offered by \xxx{} were clearly advantageous.

\begin{table}[htbp]
\vspace{5pt}
\setlength\abovecaptionskip{6pt}
\setlength\belowcaptionskip{-5pt}
    \centering
    \begin{tabular}{|c|c|c|c|}
    \hline
    &  NNTO & Cricket & \xxx \\ \hline
    
    RPCs for each inference&  NA & 5895 & 11\\ \hline
    \makecell{Average GPU utilization \\ on the GPU server }&       29.0\%  &  1.1\%  & 27.5\%   \\ \hline
    \end{tabular}
\caption{ \small \textbf{Comparison between \xxx{} and the baselines about numbers of RPC calls and average GPU utilization on GPU server.}}
  \label{tab:Micro}
\end{table}

\subsection{Validation on A Wider Range of Models}

Next, we conducted a comprehensive evaluation of \xxx{} and other baseline systems across a diverse set of models commonly used in mobile devices, varying in parameter counts, as detailed in Fig.~\ref{fig:torchvision}. 
We selected the two most prevalent models for each of the three fundamental robotic tasks (object classification, semantic segmentation, and object detection) to assess the generalizability of \xxx's performance. 
Our findings confirmed that \xxx{} achieves a performance comparable to NNTO without requiring any modifications to the source code. 
Cricket sometimes exhibited slower indoors performance compared to outdoors due to its exceptionally prolonged inference times and unstable network fluctuations, as shown in Fig.~\ref{fig:bandwidth}.
Although \xxx{} consistently outperforms in various models, performance gains are relatively smaller for models with fewer parameters.
This observation can be attributed to the fact that models with larger computational demands benefit more significantly from the robust computing power of the GPU server. 
Additionally, the substantial energy consumption incurred by extensive computations on the robot suggests that models with a larger number of parameters are more suited for offloading, thus deriving greater benefits from offloading.
\revison{It is also pertinent to note that our experimental robot (Fig.~\ref{fig:device-inference}) possesses considerably higher computational power than typical mobile devices (e.g., smartphones), suggesting \xxx{}'s benefits could be even more pronounced on more resource-constrained mobile devices.}


\section{Related Work and Discussion}
\label{sec:discussion}
\myparagraph{Fixed Calculation Logic}
\xxx{} leverages the characteristic that operators in the inference of a DNN model often follow a fixed order, allowing its record/replay mechanism to support other computational tasks~\cite{chowdhury2022optimal}, not solely SAMs, provided they exhibit fixed computational logic. 
However, \xxx{} is unable to support tasks with unfixed computational logic, such as DAMs with changing operator sequences or tasks involving complex logic and branching, due to its inability to replay varying operator sequences. 
Moreover, tasks that entail complex logic and branching are generally more suited for CPU rather than GPU execution~\cite{rosenfeld2022query}, and optimizing inference for DAMs with changing operator sequences remains a pervasive challenge across all offloading systems.

\revison{
\myparagraph{Layer partitioning}
Layer partitioning techniques~\cite{kang2017neurosurgeon} optimize individual inference requests by distributing ML model layers across mobile devices and GPU servers, thereby enhancing end-to-end inference speed and energy efficiency while addressing the transmission failure and network bandwidth fluctuation of wireless links; their widespread adoption in mobile applications has consequently drawn significant research attention, as optimal strategies depend on factors such as model architecture and application-specific trade-offs between these performance metrics. 
Historically, such layer partitioning has predominantly favored non-transparent offloading systems, largely due to their prioritization of maximal end-to-end performance and minimal transmission overhead over implementation transparency, coupled with the inherent challenge for transparent offloading systems to acquire the model architecture information essential for effective partitioning. 
However, \xxx{} offers a path to integrate these established methods within a transparent framework by directly addressing these historical hurdles. 
Its capability to achieve performance comparable to non-transparent systems alleviates concerns regarding performance prioritization. 
Furthermore, \xxx{} overcomes the architectural information barrier by utilizing data dependencies within the inference operator sequence, identifiable through its operator sequence search, to discern the model architecture. 
This understanding subsequently allows for the adaptation of layer partitioning strategies to \xxx{}, refining scheduling granularity from the layer level to the more fine-grained operator level, all while maintaining transparency.}

\revison{
\myparagraph{Multiple inference Scheduling}
It schedules multiple DNN inference requests to optimize overall latency and energy consumption while leveraging existing layer partitioning approaches for individual executions. 
Various decision algorithms (e.g., urgency-based prioritization~\cite{lin2019computation}, deep reinforcement learning-based control~\cite{fang2017qos}, and joint optimization of relay selection~\cite{zhao2019novel}) are employed to determine execution locations and timing.
Crucially, the decision algorithms inherent in such multiple inference scheduling frameworks can be seamlessly integrated into \xxx{} to determine the execution location and start time of operators within each inference task during its replay process, mirroring their established functionality in non-transparent systems. 
Consequently, by adopting these proven decision algorithms, \xxx{} can efficiently support multiple concurrent inference tasks, replicating the high performance already demonstrated by these scheduling methods in non-transparent offloading environments while avoiding the source code modification for each application.}


\myparagraph{Model Compression}
Quantization and model distillation are two most common model compression techniques for mobile devices. 
Quantization~\cite{gong2020vecq} reduces the precision of model weights and activations to lower computation costs, while model distillation~\cite{wang2021knowledge} trains a smaller model to mimic a larger one with fewer resources.
Unlike model compression techniques that trade accuracy for efficiency, offloading methods are orthogonal, achieving fast inference without compromising accuracy by effectively scheduling computational tasks.

\section{Conclusion}
\label{sec:conclusion}
In this paper, we have proposed \xxx, a high-performance transparent offloading system optimized for ML model inference in MEC. 
\xxx{} effectively alleviates communication overhead, a common bottleneck in traditional transparent offloading systems stemming from frequent RPCs, by employing a novel record/replay mechanism.
This innovative approach enables \xxx{} to achieve performance comparable to established non-transparent offloading methods, without necessitating any modifications to the application source code. 
Consequently, \xxx{} is poised to significantly advance the deployment of diverse, ML-driven mobile applications in real-world environments. 
By providing fast, energy-efficient, and seamlessly integrated inference capabilities, \xxx{} empowers mobile device systems to execute complex tasks with enhanced operational efficiency and effectiveness.


\bibliographystyle{IEEEtran}
\bibliography{references}

\begin{thebibliography}{10}
\providecommand{\url}[1]{#1}
\csname url@samestyle\endcsname
\providecommand{\newblock}{\relax}
\providecommand{\bibinfo}[2]{#2}
\providecommand{\BIBentrySTDinterwordspacing}{\spaceskip=0pt\relax}
\providecommand{\BIBentryALTinterwordstretchfactor}{4}
\providecommand{\BIBentryALTinterwordspacing}{\spaceskip=\fontdimen2\font plus
\BIBentryALTinterwordstretchfactor\fontdimen3\font minus \fontdimen4\font\relax}
\providecommand{\BIBforeignlanguage}[2]{{%
\expandafter\ifx\csname l@#1\endcsname\relax
\typeout{** WARNING: IEEEtran.bst: No hyphenation pattern has been}%
\typeout{** loaded for the language `#1'. Using the pattern for}%
\typeout{** the default language instead.}%
\else
\language=\csname l@#1\endcsname
\fi
#2}}
\providecommand{\BIBdecl}{\relax}
\BIBdecl

\bibitem{luo2021binarized}
F.~Luo, S.~Khan, Y.~Huang, and K.~Wu, ``{Binarized Neural Network for Edge Intelligence of Sensor-Based Human Activity Recognition},'' \emph{{IEEE} Trans. Mobile Comput.}, vol.~22, no.~3, pp. 1356--1368, Mar. 2023.

\bibitem{tang2024merit}
Y.~Tang, Z.~Chen, A.~Li, T.~Zheng, Z.~Lin, J.~Xu, P.~Lv, Z.~Sun, and Y.~Gao, ``{MERIT: Multimodal Wearable Vital Sign Waveform Monitoring},'' \emph{arXiv preprint arXiv:2410.00392}, 2024.

\bibitem{saridena2022dnn}
A.~N. Saridena and A.~Choromanska, ``{DNN Patching: Progressive Fixing and Augmenting the Functionalities of DNNs for Autonomous Vehicles},'' \emph{{IEEE} Robot. Autom. Lett.}, vol.~7, no.~2, pp. 3257--3264, Apr. 2022.

\bibitem{lin2022channel}
Z.~Lin, L.~Wang, J.~Ding, B.~Tan, and S.~Jin, ``{Channel Power Gain Estimation for Terahertz Vehicle-to-Infrastructure Networks},'' \emph{{IEEE} Commun. Lett.}, vol.~27, no.~1, pp. 155--159, 2022.

\bibitem{fang2024ic3m}
Z.~Fang, Z.~Lin, S.~Hu, H.~Cao, Y.~Deng, X.~Chen, and Y.~Fang, ``{IC3M: In-Car Multimodal Multi-Object Monitoring for Abnormal Status of Both Driver and Passengers},'' \emph{arXiv preprint arXiv:2410.02592}, 2024.

\bibitem{lin2024fedsn}
Z.~Lin, Z.~Chen, Z.~Fang, X.~Chen, X.~Wang, and Y.~Gao, ``Fedsn: A federated learning framework over heterogeneous leo satellite networks,'' \emph{IEEE Transactions on Mobile Computing}, 2024.

\bibitem{zhao2019novel}
Z.~Zhao, R.~Zhao, J.~Xia, X.~Lei, D.~Li, C.~Yuen, and L.~Fan, ``{A Novel Framework of Three-Hierarchical Offloading Optimization for MEC in Industrial IoT Networks},'' \emph{{IEEE} Trans. Ind. Informat.}, vol.~16, no.~8, pp. 5424--5434, 2019.

\bibitem{peng2025sigchord}
J.~Peng, J.~Duan, Z.~Lin, H.~Yuan, Y.~Gao, and Z.~Chen, ``{SigChord: Sniffing Wide Non-Sparse Multiband Signals for Terrestrial and Non-Terrestrial Wireless Networks},'' \emph{arXiv preprint arXiv:2504.06587}, 2025.

\bibitem{yuan2025constructing}
H.~Yuan, Z.~Chen, Z.~Lin, J.~Peng, Y.~Zhong, X.~Hu, S.~Xue, W.~Li, and Y.~Gao, ``{Constructing 4D Radio Map in LEO Satellite Networks with Limited Samples},'' \emph{{IEEE} INFOCOM}, 2025.

\bibitem{chen2021rf}
Z.~Chen, C.~Cai, T.~Zheng, J.~Luo, J.~Xiong, and X.~Wang, ``{RF-Based Human Activity Recognition Using Signal Adapted Convolutional Neural Network},'' \emph{{IEEE} Trans. Mobile Comput.}, vol.~22, no.~1, pp. 487--499, 2021.

\bibitem{lin2021spatial}
Z.~Lin, L.~Wang, B.~Tan, and X.~Li, ``Spatial-spectral terahertz networks,'' \emph{IEEE Transactions on Wireless Communications}, vol.~21, no.~6, pp. 3881--3892, 2021.

\bibitem{zhang2024fedac}
Y.~Zhang, H.~Chen, Z.~Lin, Z.~Chen, and J.~Zhao, ``Fedac: An adaptive clustered federated learning framework for heterogeneous data,'' \emph{arXiv preprint arXiv:2403.16460}, 2024.

\bibitem{yuan2024satsense}
H.~Yuan, Z.~Chen, Z.~Lin, J.~Peng, Z.~Fang, Y.~Zhong, Z.~Song, and Y.~Gao, ``{SatSense: Multi-Satellite Collaborative Framework for Spectrum Sensing},'' \emph{{IEEE} Trans. Cogn. Commun. Netw.}, 2025.

\bibitem{peng2024sums}
J.~Peng, Z.~Chen, Z.~Lin, H.~Yuan, Z.~Fang, L.~Bao, Z.~Song, Y.~Li, J.~Ren, and Y.~Gao, ``{SUMS: Sniffing Unknown Multiband Signals under Low Sampling Rates},'' \emph{{IEEE} Trans. Mobile Comput.}, 2024.

\bibitem{zhao2024leo}
Z.~Zhao, Z.~Chen, Z.~Lin, W.~Zhu, K.~Qiu, C.~You, and Y.~Gao, ``{LEO Satellite Networks Assisted Geo-Distributed Data Processing},'' \emph{{IEEE} Wireless Commun. Lett.}, 2024.

\bibitem{lin2025hierarchical}
Z.~Lin, W.~Wei, Z.~Chen, C.-T. Lam, X.~Chen, Y.~Gao, and J.~Luo, ``{Hierarchical Split Federated Learning: Convergence Analysis and System Optimization},'' \emph{{IEEE} Trans. Mobile Comput.}, 2025.

\bibitem{kapao}
W.~McNally, K.~Vats, A.~Wong, and J.~McPhee, ``{Rethinking Keypoint Representations: Modeling Keypoints and Poses as Objects for Multi-Person Human Pose Estimation},'' in \emph{{Proc. ECCV}}, Oct. 2022, pp. 37--54.

\bibitem{lin2024efficient}
Z.~Lin, G.~Zhu, Y.~Deng, X.~Chen, Y.~Gao, K.~Huang, and Y.~Fang, ``{Efficient Parallel Split Learning over Resource-Constrained Wireless Edge Networks},'' \emph{{IEEE} Trans. Mobile Comput.}, vol.~23, no.~10, pp. 9224--9239, 2024.

\bibitem{agrnav}
J.~Wang, Z.~Sun, X.~Guan, T.~Shen, Z.~Zhang, T.~Duan, D.~Huang, S.~Zhao, and H.~Cui, ``{AGRNav: Efficient and Energy-Saving Autonomous Navigation for Air-Ground Robots in Occlusion-Prone Environments},'' in \emph{{Proc. ICRA}}, May 2024, pp. 4494--4501.

\bibitem{duan2025rethinking}
T.~Duan, Z.~Zhang, Z.~Lin, Y.~Gao, L.~Xiong, Y.~Cui, H.~Liang, X.~Chen, H.~Cui, and D.~Huang, ``Rethinking adversarial attacks in reinforcement learning from policy distribution perspective,'' in \emph{ICASSP 2025-2025 IEEE International Conference on Acoustics, Speech and Signal Processing (ICASSP)}, 2025, pp. 1--5.

\bibitem{lin2023pushing}
Z.~Lin, G.~Qu, Q.~Chen, X.~Chen, Z.~Chen, and K.~Huang, ``{Pushing Large Language Models to the 6G Edge: Vision, Challenges, and Opportunities},'' \emph{arXiv preprint arXiv:2309.16739}, 2023.

\bibitem{zhang2025state}
Z.~Zhang, T.~Duan, Z.~Lin, D.~Huang, Z.~Fang, Z.~Sun, L.~Xiong, H.~Liang, H.~Cui, and Y.~Cui, ``State-aware perturbation optimization for robust deep reinforcement learning,'' \emph{arXiv preprint arXiv:2503.20613}, 2025.

\bibitem{lin2025hasfl}
Z.~Lin, Z.~Chen, X.~Chen, W.~Ni, and Y.~Gao, ``{HASFL: Heterogeneity-Aware Split Federated Learning over Edge Computing Systems},'' \emph{arXiv preprint arXiv:2506.08426}, 2025.

\bibitem{cao2022monoscene}
A.-Q. Cao and R.~de~Charette, ``{MonoScene: Monocular 3D Semantic Scene Completion},'' in \emph{{Proc. CVPR}}, Jun. 2022, pp. 3991--4001.

\bibitem{lyu2023optimal}
S.~Lyu, Z.~Lin, G.~Qu, X.~Chen, X.~Huang, and P.~Li, ``Optimal resource allocation for u-shaped parallel split learning,'' in \emph{2023 IEEE Globecom Workshops (GC Wkshps)}, 2023, pp. 197--202.

\bibitem{lin2024splitlora}
Z.~Lin, X.~Hu, Y.~Zhang, Z.~Chen, Z.~Fang, X.~Chen, A.~Li, P.~Vepakomma, and Y.~Gao, ``{SplitLoRA: A Split Parameter-Efficient Fine-Tuning Framework for Large Language Models},'' \emph{arXiv preprint arXiv:2407.00952}, 2024.

\bibitem{sun2025intra}
Z.~Sun, X.~Guan, Z.~Lin, Z.~Fang, X.~Cai, Z.~Chen, F.~Liu, H.~Cui, J.~Xiong, W.~Ni \emph{et~al.}, ``Intra-dp: A high performance collaborative inference system for mobile edge computing,'' \emph{arXiv preprint arXiv:2507.05829}, 2025.

\bibitem{lin2024adaptsfl}
Z.~Lin, G.~Qu, W.~Wei, X.~Chen, and K.~K. Leung, ``{Adaptsfl: Adaptive Split Federated Learning in Resource-Constrained Edge Networks},'' \emph{{IEEE} Trans. Netw.}, 2024.

\bibitem{abbas2017mobile}
N.~Abbas, Y.~Zhang, A.~Taherkordi, and T.~Skeie, ``{Mobile Edge Computing: A Survey},'' \emph{{IEEE} Internet Things J.}, vol.~5, no.~1, pp. 450--465, Feb. 2018.

\bibitem{qiao2020best}
B.~Qiao, M.~A. {\"O}zkan, J.~Teich, and F.~Hannig, ``{The Best of Both Worlds: Combining CUDA Graph with an Image Processing DSL},'' in \emph{{Proc. DAC}}, 2020, pp. 1--6.

\bibitem{cricket}
N.~Eiling, J.~Baude, S.~Lankes, and A.~Monti, ``{Cricket: A Virtualization Layer for Distributed Execution of CUDA Applications with Checkpoint/Restart Support},'' \emph{{Concurrency Comput. Pract. Exp.}}, vol.~34, no.~14, p. e6474, May 2022.

\bibitem{davoodi2019tensorrt}
P.~Davoodi, C.~Gwon, G.~Lai, and T.~Morris, ``{TensorRT Inference with {TensorFlow}},'' in \emph{Proc. {GPU} Technol. Conf. ({GTC})}, Mar. 2019.

\bibitem{yi2024study}
X.~Yi, ``{A Study of Performance Programming of CPU, GPU accelerated Computers and SIMD Architecture},'' \emph{arXiv preprint arXiv:2409.10661}, 2024.

\bibitem{mounesan2025infer}
M.~Mounesan, X.~Zhang, and S.~Debroy, ``{Infer-EDGE: Dynamic DNN Inference Optimization in `Just-in-Time' Edge-AI Implementations},'' \emph{arXiv preprint arXiv:2501.18842}, 2025.

\bibitem{devito2022torchscript}
Z.~DeVito, ``{TorchScript: Optimized Execution of PyTorch Programs},'' \emph{{Retrieved January}}, 2022.

\bibitem{pytorch}
A.~Paszke, S.~Gross, F.~Massa, A.~Lerer, J.~Bradbury, G.~Chanan, T.~Killeen, Z.~Lin, N.~Gimelshein, L.~Antiga, A.~Desmaison, A.~Kopf, E.~Yang, Z.~DeVito, M.~Raison, A.~Tejani, S.~Chilamkurthy, B.~Steiner, L.~Fang, J.~Bai, and S.~Chintala, ``{PyTorch: An Imperative Style, High-Performance Deep Learning Library},'' \emph{arXiv preprint arXiv:1912.01703}, Dec. 2019.

\bibitem{cudaruntime}
S.~Kato, ``{Implementing Open-Source CUDA Runtime},'' in \emph{Proc. of the 54the Programming Symposium}, Hakone, Japan, Jan. 2013, pp. 111--118.

\bibitem{schneider2022evaluation}
M.~Schneider, F.~Haag, A.~K. Khalil, and D.~A. Breunig, ``{Evaluation of Communication Technologies for Distributed Industrial Control Systems: Concept and Evaluation of 5G and WiFi 6},'' \emph{{Procedia CIRP}}, vol. 107, pp. 588--593, 2022.

\bibitem{tensorflow}
M.~Abadi, P.~Barham, J.~Chen, Z.~Chen, A.~Davis, J.~Dean, M.~Devin, S.~Ghemawat, G.~Irving, M.~Isard \emph{et~al.}, ``{TensorFlow: A System for Large-Scale Machine Learning},'' in \emph{{Proc. OSDI}}, Nov. 2016, pp. 265--283.

\bibitem{opencl}
A.~Munshi, ``{The {OpenCL} Specification},'' in \emph{Proc. IEEE Hot Chips 21 Symp. (HCS)}, Aug. 2009.

\bibitem{jia2022codl}
F.~Jia, D.~Zhang, T.~Cao, S.~Jiang, Y.~Liu, J.~Ren, and Y.~Zhang, ``{CoDL: Efficient {CPU-GPU} Co-Execution for Deep Learning Inference on Mobile Devices},'' in \emph{{Proc. MobiSys}}, Jun. 2022, pp. 209--221.

\bibitem{plancher2021accelerating}
B.~Plancher, S.~M. Neuman, T.~Bourgeat, S.~Kuindersma, S.~Devadas, and V.~J. Reddi, ``{Accelerating Robot Dynamics Gradients on a CPU, GPU, and FPGA},'' \emph{{IEEE} Robot. Autom. Lett.}, vol.~6, no.~2, pp. 2335--2342, Apr. 2021.

\bibitem{lane2016deepx}
N.~D. Lane, S.~Bhattacharya, P.~Georgiev, C.~Forlivesi, L.~Jiao, L.~Qendro, and F.~Kawsar, ``{DeepX: A Software Accelerator for Low-Power Deep Learning Inference on Mobile Devices},'' in \emph{{Proc. IPSN}}, Apr. 2016, pp. 1--12.

\bibitem{ghosh2023react}
A.~Ghosh, S.~Iyengar, S.~Lee, A.~Rathore, and V.~N. Padmanabhan, ``{REACT: Streaming Video Analytics on the Edge with Asynchronous Cloud Support},'' in \emph{{Proc. IoTDI}}, May 2023, pp. 222--235.

\bibitem{iphone}
``{MLPerf Mobile Benchmarks},'' \url{https://mlcommons.org/working-groups/benchmarks/mobile/}, 2023.

\bibitem{raspberrypi}
RaspberryPi, ``{Raspberry Pi},'' \url{https://www.raspberrypi.com/documentation/computers/configuration.html\#power-consumption}, 2022.

\bibitem{jetsonnx}
{NVIDIA}, ``{Jetson Xavier NX Series: The World's Smallest AI Supercomputer},'' \url{https://www.nvidia.com/en-us/autonomous-machines/embedded-systems/jetson-xavier-nx-developer-kit/}, 2024.

\bibitem{ning2024power}
Z.~Ning, M.~Vandersteegen, K.~Van~Beeck, T.~Goedem{\'e}, and P.~Vandewalle, ``{Power Consumption Benchmark for Embedded AI Inference},'' in \emph{Proc. Int. Conf. Appl. Comput. WWW/Internet (AC)}, Jan. 2024, pp. 3--10.

\bibitem{armanfard2015local}
N.~Armanfard, J.~P. Reilly, and M.~Komeili, ``{Local Feature Selection for Data Classification},'' \emph{{IEEE} Trans. Pattern Anal. Mach. Intell.}, vol.~38, no.~6, pp. 1217--1227, 2015.

\bibitem{kang2017neurosurgeon}
Y.~Kang, J.~Hauswald, C.~Gao, A.~Rovinski, T.~Mudge, J.~Mars, and L.~Tang, ``{Neurosurgeon: Collaborative Intelligence Between the Cloud and Mobile Edge},'' in \emph{{Proc. ASPLOS}}, Apr. 2017, pp. 615--629.

\bibitem{lin2019computation}
L.~Lin, X.~Liao, H.~Jin, and P.~Li, ``{Computation Offloading Toward Edge Computing},'' \emph{{Proc. IEEE}}, vol. 107, no.~8, pp. 1584--1607, Aug. 2019.

\bibitem{fang2017qos}
Z.~Fang, T.~Yu, O.~J. Mengshoel, and R.~K. Gupta, ``{QoS-Aware Scheduling of Heterogeneous Servers for Inference in Deep Neural Networks},'' in \emph{{Proc. CIKM}}, Nov. 2017, pp. 2067--2070.

\bibitem{infiniBand}
NVIDIA, ``{InfiniBand Networking Solutions},'' \url{https://www.nvidia.com/en-us/networking/products/infiniband/}, 2024.

\bibitem{liu2023first}
R.~Liu and N.~Choi, ``{A First Look at Wi-Fi 6 in Action: Throughput, Latency, Energy Efficiency, and Security},'' in \emph{{Proc. ACM Meas. Anal. Comput. Syst.}}, vol.~7, no.~1, Mar. 2023, pp. 1--25.

\bibitem{yang2022mobile}
X.~Yang, H.~Lin, Z.~Li, F.~Qian, X.~Li, Z.~He, X.~Wu, X.~Wang, Y.~Liu, Z.~Liao \emph{et~al.}, ``{Mobile Access Bandwidth in Practice: Measurement, Analysis, and Implications},'' in \emph{{Proc. SIGCOMM}}, Aug. 2022, pp. 114--128.

\bibitem{masiukiewicz2019throughput}
A.~Masiukiewicz, ``{Throughput Comparison between The New HEW 802.11 ax Standard and 802.11 n/ac Standards in Selected Distance Windows},'' \emph{{Int. J. Electron. Telecommun.}}, vol.~65, no.~1, pp. 79--84, 2019.

\bibitem{ding2015performance}
M.~Ding, P.~Wang, D.~L{\'o}pez-P{\'e}rez, G.~Mao, and Z.~Lin, ``{Performance Impact of LoS and NLoS Transmissions in Dense Cellular Networks},'' \emph{{IEEE} Trans. Wireless Commun.}, vol.~15, no.~3, pp. 2365--2380, Mar. 2016.

\bibitem{ren2018proportional}
Y.~Ren, C.-W. Tung, J.-C. Chen, and F.~Y. Li, ``{Proportional and Preemption-Enabled Traffic Offloading for IP Flow Mobility: Algorithms and Performance Evaluation},'' \emph{{IEEE} Trans. Veh. Technol.}, vol.~67, no.~12, pp. 12\,095--12\,108, Dec. 2018.

\bibitem{iperf}
``{{iPerf} - {Download} {iPerf3} and Original {iPerf} Pre-Compiled Binaries},'' \url{https://iperf.fr/iperf-download.php}, 2024.

\bibitem{tian2005tcp}
Y.~Tian, K.~Xu, and N.~Ansari, ``{TCP in Wireless Environments: Problems and Solutions},'' \emph{{IEEE} Commun. Mag.}, vol.~43, no.~3, pp. S27--S32, Mar. 2005.

\bibitem{nsightcompute}
{NVIDIA}, ``{NVIDIA Nsight Compute Documentation},'' \url{https://docs.nvidia.com/nsight-compute/index.html}, 2024.

\bibitem{libtirpc}
S.~Dickson, ``{libtirpc: Transport Independent RPC library},'' \url{https://git.linux-nfs.org/?p=steved/libtirpc.git}, 2024.

\bibitem{singhvi2021cliquemap}
A.~Singhvi, A.~Akella, M.~Anderson, R.~Cauble, H.~Deshmukh, D.~Gibson, M.~M. Martin, A.~Strominger, T.~F. Wenisch, and A.~Vahdat, ``{Cliquemap: Productionizing an RMA-Based Distributed Caching System},'' in \emph{{Proc. SIGCOMM}}.\hskip 1em plus 0.5em minus 0.4em\relax {ACM}, 2021, pp. 93--105.

\bibitem{lazarev2021dagger}
N.~Lazarev, S.~Xiang, N.~Adit, Z.~Zhang, and C.~Delimitrou, ``{Dagger: Efficient and Fast RPCs in Cloud Microservices With Near-Memory Reconfigurable NICs},'' in \emph{{Proc. ASPLOS}}.\hskip 1em plus 0.5em minus 0.4em\relax {ACM}, 2021, pp. 36--51.

\bibitem{eyerman2022efficient}
S.~Eyerman and I.~Hur, ``{Efficient Asynchronous RPC Calls for Microservices: DeathStarBench Study},'' \emph{arXiv preprint arXiv:2209.13265}, 2022.

\bibitem{huang2020interpretable}
Z.~Huang and Y.~Li, ``{Interpretable and Accurate Fine-Grained Recognition via Region Grouping},'' in \emph{{Proc. CVPR}}, 2020, pp. 8662--8672.

\bibitem{wang2019gmc}
H.~Wang, Y.~Yang, and B.~Liu, ``{GMC: Graph-Based Multi-View Clustering},'' \emph{{IEEE} Trans. Knowledge Data Eng.}, vol.~32, no.~6, pp. 1116--1129, 2019.

\bibitem{lewis2020retrieval}
P.~Lewis, E.~Perez, A.~Piktus, F.~Petroni, V.~Karpukhin, N.~Goyal, H.~K{\"u}ttler, M.~Lewis, W.-t. Yih, T.~Rockt{\"a}schel \emph{et~al.}, ``{Retrieval-Augmented Generation for Knowledge-Intensive NLP Tasks},'' in \emph{{Proc. NeurIPS}}, vol.~33, 2020, pp. 9459--9474.

\bibitem{devlin2018bert}
J.~Devlin, M.-W. Chang, K.~Lee, and K.~Toutanova, ``{BERT: Pre-Training of Deep Bidirectional Transformers for Language Understanding},'' \emph{arXiv preprint arXiv:1810.04805}, Oct. 2018.

\bibitem{bai2018empirical}
S.~Bai, J.~Z. Kolter, and V.~Koltun, ``{An Empirical Evaluation of Generic Convolutional and Recurrent Networks for Sequence Modeling},'' \emph{arXiv preprint arXiv:1803.01271}, 2018.

\bibitem{dosovitskiy2020image}
A.~Dosovitskiy, L.~Beyer, A.~Kolesnikov, D.~Weissenborn, X.~Zhai, T.~Unterthiner, M.~Dehghani, M.~Minderer, G.~Heigold, S.~Gelly \emph{et~al.}, ``{An Image Is Worth 16x16 Words: Transformers for Image Recognition at Scale},'' in \emph{{Proc. ICLR}}, 2020.

\bibitem{brown2020language}
T.~Brown, B.~Mann, N.~Ryder, M.~Subbiah, J.~D. Kaplan, P.~Dhariwal, A.~Neelakantan, P.~Shyam, G.~Sastry, A.~Askell \emph{et~al.}, ``{Language Models Are Few-Shot Learners},'' in \emph{{Proc. NeurIPS}}, vol.~33, 2020, pp. 1877--1901.

\bibitem{shazeer2017outrageously}
N.~Shazeer, A.~Mirhoseini, K.~Maziarz, A.~Davis, Q.~Le, G.~Hinton, and J.~Dean, ``{Outrageously Large Neural Networks: The Sparsely-Gated Mixture-of-Experts Layer},'' in \emph{{Proc. ICLR}}, 2017.

\bibitem{sutskever2014sequence}
I.~Sutskever, O.~Vinyals, and Q.~V. Le, ``{Sequence to Sequence Learning with Neural Networks},'' in \emph{{Proc. NeurIPS}}, vol.~27, 2014.

\bibitem{chen2016xgboost}
T.~Chen and C.~Guestrin, ``{XGBoost: A Scalable Tree Boosting System},'' in \emph{{Proc. KDD}}, 2016, pp. 785--794.

\bibitem{jiang2019semi}
B.~Jiang, Z.~Zhang, D.~Lin, J.~Tang, and B.~Luo, ``{Semi-Supervised Learning with Graph Learning-Convolutional Networks},'' in \emph{{Proc. CVPR}}, 2019, pp. 11\,313--11\,320.

\bibitem{tarnawski2020efficient}
J.~M. Tarnawski, A.~Phanishayee, N.~Devanur, D.~Mahajan, and F.~N. Paravecino, ``{Efficient Algorithms for Device Placement of DNN Graph Operators},'' in \emph{{Proc. NeurIPS}}, H.~Larochelle, M.~Ranzato, R.~Hadsell, M.~F. Balcan, and H.~Lin, Eds., vol.~33, Dec. 2020, pp. 15\,451--15\,463.

\bibitem{charalampopoulos2021faster}
P.~Charalampopoulos, T.~Kociumaka, S.~P. Pissis, and J.~Radoszewski, ``{Faster Algorithms for Longest Common Substring},'' \emph{arXiv preprint arXiv:2105.03106}, 2021.

\bibitem{torchvision}
S.~Marcel and Y.~Rodriguez, ``{Torchvision the Machine-Vision Package of Torch},'' in \emph{{Proc. ACM Multimedia}}, Oct. 2010, pp. 1485--1488.

\bibitem{he2015deepresiduallearningimage}
K.~He, X.~Zhang, S.~Ren, and J.~Sun, ``{Deep Residual Learning for Image Recognition},'' \emph{arXiv preprint arXiv:1512.03385}, Dec. 2015.

\bibitem{liu2022convnet2020s}
Z.~Liu, H.~Mao, C.-Y. Wu, C.~Feichtenhofer, T.~Darrell, and S.~Xie, ``{A ConvNet for the 2020s},'' \emph{arXiv preprint arXiv:2201.03545}, Jan. 2022.

\bibitem{long2015fullyconvolutionalnetworkssemantic}
J.~Long, E.~Shelhamer, and T.~Darrell, ``{Fully Convolutional Networks for Semantic Segmentation},'' \emph{arXiv preprint arXiv:1411.4038}, Nov. 2015.

\bibitem{chen2017rethinkingatrousconvolutionsemantic}
L.-C. Chen, G.~Papandreou, F.~Schroff, and H.~Adam, ``{Rethinking Atrous Convolution for Semantic Image Segmentation},'' \emph{arXiv preprint arXiv:1706.05587}, Jun. 2017.

\bibitem{ren2016fasterrcnnrealtimeobject}
S.~Ren, K.~He, R.~Girshick, and J.~Sun, ``{Faster R-CNN: Towards Real-Time Object Detection with Region Proposal Networks},'' \emph{arXiv preprint arXiv:1506.01497}, Jun. 2015.

\bibitem{lin2018focallossdenseobject}
T.-Y. Lin, P.~Goyal, R.~Girshick, K.~He, and P.~Doll{\'a}r, ``{Focal Loss for Dense Object Detection},'' \emph{arXiv preprint arXiv:1708.02002}, Aug. 2017.

\bibitem{noormohammadpour2017datacenter}
M.~Noormohammadpour and C.~S. Raghavendra, ``{Datacenter Traffic Control: Understanding Techniques and Tradeoffs},'' \emph{{IEEE} Commun. Surv. Tutor.}, vol.~20, no.~2, pp. 1492--1525, Jun. 2018.

\bibitem{glenn_jocher_2022_7347926}
G.~Jocher, A.~Chaurasia, A.~Stoken, J.~Borovec, NanoCode012, Y.~Kwon, K.~Michael, T.~Xie, J.~Fang, imyhxy, Lorna, Z.~Yifu, C.~Wong, A.~V, D.~Montes, Z.~Wang, C.~Fati, J.~Nadar, Laughing, UnglvKitDe, V.~Sonck, tkianai, yxNONG, P.~Skalski, A.~Hogan, D.~Nair, M.~Strobel, and M.~Jain, ``{ultralytics/yolov5: v7.0 - YOLOv5 SOTA Realtime Instance Segmentation},'' Nov. 2022.

\bibitem{corporation_pynvml_nodate}
{NVIDIA}, ``{pynvml: {Python} {Bindings} for the {NVIDIA} {Management} {Library}},'' \url{https://developer.nvidia.com/management-library-nvml/}, 2024.

\bibitem{chowdhury2022optimal}
R.~Chowdhury and D.~Subramani, ``{Optimal Path Planning of Autonomous Marine Vehicles in Stochastic Dynamic Ocean Flows Using a GPU-Accelerated Algorithm},'' \emph{{IEEE J. Ocean. Eng.}}, vol.~47, no.~4, pp. 864--879, Oct. 2022.

\bibitem{rosenfeld2022query}
V.~Rosenfeld, S.~Bre{\ss}, and V.~Markl, ``{Query Processing on Heterogeneous CPU/GPU Systems},'' \emph{{ACM Comput. Surv.}}, vol.~55, no.~1, pp. 1--38, Jan. 2023.

\bibitem{gong2020vecq}
C.~Gong, Y.~Chen, Y.~Lu, T.~Li, C.~Hao, and D.~Chen, ``{VecQ: Minimal Loss DNN Model Compression with Vectorized Weight Quantization},'' \emph{{IEEE} Trans. Comput.}, vol.~70, no.~5, pp. 696--710, May 2021.

\bibitem{wang2021knowledge}
L.~Wang and K.-J. Yoon, ``{Knowledge Distillation and Student-Teacher Learning for Visual Intelligence: A Review and New Outlooks},'' \emph{{IEEE} Trans. Pattern Anal. Mach. Intell.}, vol.~44, no.~6, pp. 3048--3068, Jun. 2022.

\end{thebibliography}

\end{document}